\documentclass[11pt]{article}
\usepackage[utf8]{inputenc}
\usepackage{amsmath}
\usepackage{amsfonts}
\usepackage{amssymb}
\usepackage{amsthm}

\usepackage{graphicx}
\usepackage{graphicx}
\usepackage[left=2cm,right=2cm,top=3cm,bottom=3cm]{geometry}
\usepackage{caption}
\usepackage{multirow}
\newtheorem{theorem}{Theorem}

\newtheorem{definition}[theorem]{Definition}
\newtheorem{proposition}[theorem]{Proposition}

\title{\textbf{Robust copula estimation for one-shot devices with correlated failure modes}}
\date{}
\author{E. Castilla\footnote{Universidad Rey Juan Carlos, Spain. elena.castilla@urjc.es} \ \ and \ P.J. Chocano\footnote{Universidad Rey Juan Carlos, Spain. pedro.chocano@urjc.es}}
\begin{document}
\maketitle

    \captionsetup{width=0.9\textwidth}

\abstract{This paper presents a robust method for estimating copula models to evaluate the dependence between failure modes in one-shot devices—systems designed for single use and destroyed upon activation. Traditional approaches, such as maximum likelihood estimation (MLE), often produce unreliable results when faced with outliers or model misspecification. To address  these limitations, we introduce a divergence-based estimation technique that enhances robustness and provides a more reliable characterization of the joint failure-time distribution. Extensive simulation studies confirm the robustness of the proposed method. Additionally, we illustrate its practical utility through the analysis of a real-world dataset.}

\section{Introduction}
\label{sec:1}
One-shot devices play a critical role in various high-stakes applications, including automobile airbags, military systems, space missions, and battery technologies. These devices are designed for a single use and are destroyed upon activation, making traditional reliability assessment particularly challenging. Since lifetime data are typically unavailable due to the device's destruction, reliability studies for one-shot devices must rely on censored observations—specifically, left- or right-censored data—depending on whether the failure occurs before or after the inspection time. For a comprehensive overview of one-shot device testing, see the books by Balakrishnan et al. \cite{Balakrishnan2021} and Balakrishnan and Castilla \cite{Balakrishnan2025}.

Most prior work on one-shot device reliability has focused on scenarios involving a single failure mode. In these scenarios, the status of each item is binary, indicating whether its lifetime exceeds the inspection time, without distinguishing between different failure modes. Research in this area has primarily concentrated on parameter estimation using parametric approaches. 
Early studies concentrated on constant-stress accelerated life tests (CSALTs), in which all devices are subjected to elevated stress levels to induce failures over a relatively short period. For example, \cite{Balakrishnan2012} developed an expectation–maximization (EM) algorithm for maximum likelihood estimation under the assumption of exponential lifetimes. This methodology was subsequently extended to Weibull, gamma, and lognormal lifetime distributions by \cite{Balakrishnan2013}, \cite{Balakrishnan2014}, and \cite{Balakrishnan2022}, respectively. Other studies have addressed one-shot devices under step-stress accelerated life tests (SSALTs) \cite{Ling2019,Tung2025,Ling2026}, as well as under cyclic accelerated life tests \cite{Zhu2021,Zhu2022,Zhang2025}.

While these models have provided valuable insights, they fail to capture more realistic scenarios involving multiple coexisting failure modes. In practice, multiple competing failure mechanisms often contribute to device failure, complicating the assessment of system reliability. A commonly adopted approach for modeling such scenarios is the competing risks model, which assumes each failure mode follows an independent failure process. Applications of this framework to one-shot devices can be found in \cite{Balakrishnan2015a,Balakrishnan2015b,Balakrishnan2024}. However, the assumption of independence between failure modes—though mathematically convenient—is often unrealistic. Correlation between failure modes is not only common, but can significantly bias reliability estimates if ignored.

To address this limitation, copula models offer a flexible and powerful means of modeling dependence between failure mechanisms. By decoupling the marginal behavior of individual failure modes from their joint dependence structure, copulas provide a more realistic framework for capturing complex interdependencies. Copula functions have found widespread application across various fields—such as reliability, stochastic orders, and finance \cite{OJ2024,Capaldo2024,Capaldo2025a, Capaldo2025b}—due to their ability to model dependence structures beyond simple linear correlation.

Copula-based approaches have recently been applied to one-shot device data. Ling et al. \cite{Ling2021} were the first to apply copula models in this context. Using a parametric framework, they employed Archimedean copulas to analyze devices with two failure modes. Subsequently, Ashkamini and Rezaei \cite{Ashkamini2023} and Salem and Elsawah \cite{Salem2024} developed Bayesian approaches for this purpose. More recently, Prajapati et al. \cite{Prajapati2023} investigated the issue of copula model misspecification. However, all these studies have primarily relied on maximum likelihood estimation (MLE). Although MLE is known for its efficiency under ideal conditions, it is particularly sensitive to outliers and model misspecification—factors that are often present in high-stakes testing environments.

In this paper, we introduce a divergence-based estimation method for copula models that enhances robustness without sacrificing efficiency. Unlike MLE, divergence-based methods may be less affected by anomalies or deviations from model assumptions. Although divergence methods have previously been applied to one-shot testing data (see \cite{Balakrishnan2025,Balakrishnan2024,Balakrishnan2019ieee,Balakrishnan2023, Baghel2024}), their application in the context of dependent failure modes via copulas has not yet been explored.

The remainder of this paper is organized as follows. In Section \ref{sec:dataModel}, we describe the data structure of one-shot device testing with multiple failure modes and provide illustrative examples. Section \ref{sec:Copulas} reviews the fundamentals of copula models, with a focus on Archimedean copulas—specifically, the Gumbel-Hougaard and Frank families. Section \ref{sec:Estimation} introduces the proposed estimation framework and details the statistical methodology. In Section \ref{sec:numerical}, we assess the performance of the proposed method through simulation studies and demonstrate its application to two real-world datasets. Finally, Section \ref{sec:conclusions} summarizes our findings and discusses avenues for future research. We add some relevant information in the Appendix.

\section{One-shot devices with multiple failure modes \label{sec:dataModel}}

\begin{figure}[h!]
    \centering
    \includegraphics[width=0.75\linewidth]{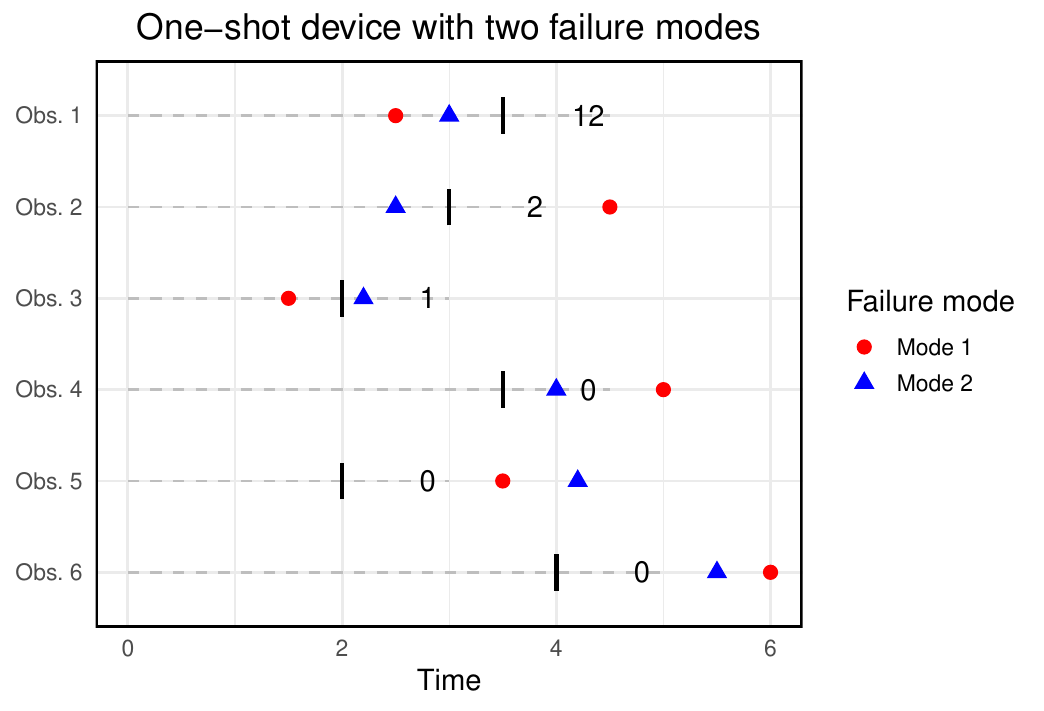}
    \caption{Schematic illustration of the observation scheme with two failure modes. For each unit, two latent failure times are compared with the inspection time, leading to four possible outcomes: no failure (0), failure in mode 1 (1), failure in mode 2 (2), and failure in both modes (12).}
    \label{fig:OneShot}
\end{figure}

We consider an experiment in which $K$ devices are subjected to a constant-stress accelerated life test (CSALT). The test is conducted at $I$ distinct inspection times and under $J$ different stress conditions. At each inspection time $IT_i$ ($i = 1, \dots, I$) and stress condition $x_j$ ($j = 1, \dots, J$), a total of $K_{ij}$ units are tested. The total number of units tested across all inspection times and stress conditions is $K = \sum_{i=1}^I \sum_{j=1}^J K_{ij}$. During the experiment, we observe the occurrence of failures under two distinct failure modes, as illustrated in Figure~\ref{fig:OneShot}. Specifically, for each combination of inspection time and stress condition, the number of devices that experience no failure is denoted by $n_{ij,0}$, the number that fail only in mode 1 is $n_{ij,1}$, the number that fail only in mode 2 is $n_{ij,2}$, and the number that fail in both modes is $n_{ij,12}$. The total number of failures corresponding to each failure mode, denoted as $N_{ij,r}$, is also recorded. This experimental design is summarized in Table \ref{tab:COPULA_model} for the specific inspection time $IT_i$ and stress condition $x_j$.

\begin{table}[h!!]
  \setlength\tabcolsep{5pt}\def\arraystretch{1.2}
  \caption{One-shot device testing under two dependent failure modes at inspection time $IT_i$ and stress level $\boldsymbol{x}_j$. \label{tab:COPULA_model}}
  \center
  \small
  \begin{tabular}{|l|ccc|}
  \hline
  & \multicolumn{3}{c|}{\textbf{Failure mode 1}} \\
  \cline{2-4}
  \textbf{Failure mode 2} & Present & Absent & Total \\ \hline
  Present & $n_{ij,12}$ & $n_{ij,2}$  & $N_{ij,2}$ \\
  Absent  & $n_{ij,1}$  & $n_{ij,0}$  & $K_{ij} - N_{ij,2}$ \\
 Total  & $N_{ij,1}$  & $K_{ij} - N_{ij,1}$  & $K_{ij}$ \\ \hline
  \end{tabular}
\end{table}

\subsection{Serial sacrifice data \label{sec:SSdata}}

This experiment, described in \cite{Berlin1979}, investigates the onset and progression of radiation-induced diseases in laboratory mice. Specifically, we focus on two broad disease categories: (I) thymic lymphoma and/or glomerulosclerosis, and (II) all other diseases.

Thymic lymphoma is a type of cancer originating in the thymus, a central organ of the immune system. It is commonly used as a marker for radiation-induced malignancy in mice due to its high incidence following exposure to ionizing radiation. Glomerulosclerosis refers to scarring or hardening of the glomeruli in the kidneys, often indicative of chronic kidney disease and a marker of systemic organ damage. The inclusion of both conditions under a single risk category reflects their biological relevance as hallmark outcomes of radiation exposure—representing oncogenic and degenerative processes, respectively.

\begin{figure}[h]
    \centering
    \includegraphics[width=0.5\linewidth]{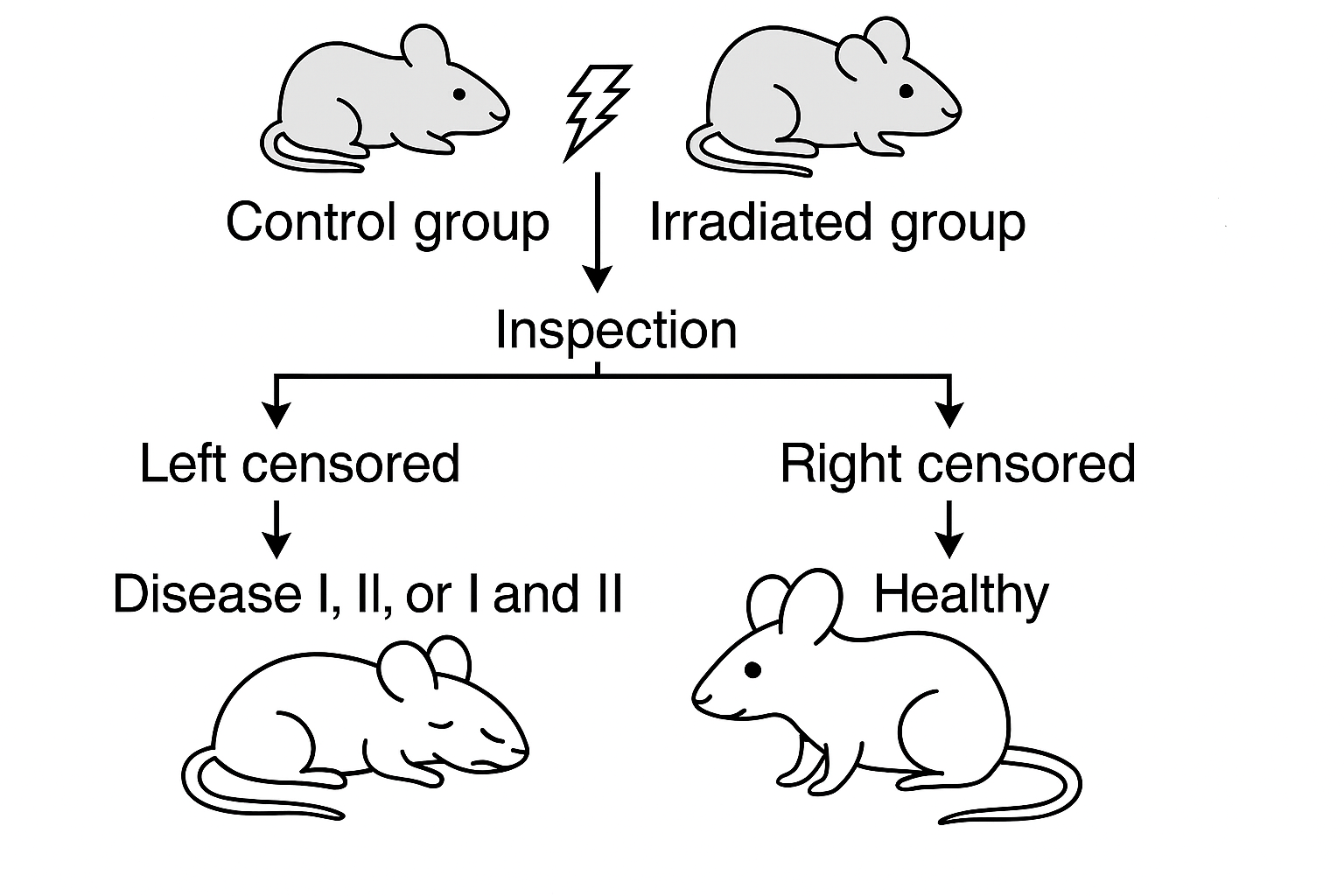}
    \caption{Schematic of the serial sacrifice experiment in mice. At scheduled times, mice are euthanized and assessed for disease status: healthy, affected by category (I), category (II), or both. Disease onset is left-censored if it occurred before inspection, and right-censored if no disease was observed.}
    \label{fig:mice}
\end{figure}

Data were collected from two groups of female mice: a control group (361 mice not exposed to radiation) and an irradiated group (343 mice exposed to $\gamma$-radiation). The mice were euthanized at scheduled inspection times—a procedure known as serial sacrifice. The onset times of diseases are subject to censoring: they are left-censored if the disease developed before the time of sacrifice and right-censored if it had not yet occurred by that time (see Figure \ref{fig:mice}). Therefore, this dataset is a suitable example of one-shot devices with multiple failure modes. Table \ref{table:SerialSacrifice} summarizes the distribution of health outcomes across the two disease categories. Moreover, since the occurrence of a disease in one category may reflect systemic biological processes that influence vulnerability to other conditions, it is reasonable to consider potential dependence between these two competing risk modes.

\begin{table}[h]\setlength\tabcolsep{5pt}\def\arraystretch{1.2}
\centering
\caption{Serial sacrifice data showing presence or absence of two disease categories (I) and (II) in mice. \label{table:SerialSacrifice}}

\begin{minipage}[t]{0.49\textwidth}
\small
\centering
\textbf{Control group}\\[6pt]
\begin{tabular}{|c| r r r r|}
\hline
Time   & \multicolumn{4}{c|}{Mice status} \\ \cline{2-5} 
 (days)  & Healthy & (I) & (II)  & (I) \& (II) \\
\hline
100 & 58 & 13 & 0 & 1 \\
200 & 40 & 23 & 1 & 1 \\
300 & 18 & 41 & 1 & 3 \\
400 & 8  & 25 & 1 & 6 \\
500 & 1  & 21 & 1 & 16 \\
600 & 1  & 11 & 0 & 21 \\
700 & 0  & 9  & 1 & 39 \\
\hline
\end{tabular}
\end{minipage}
\hfill
\begin{minipage}[t]{0.49\textwidth}
\small
\centering
\textbf{Irradiated group}\\[6pt]
\begin{tabular}{|c| r r r r|}
\hline
Time   & \multicolumn{4}{c|}{Mice status} \\ \cline{2-5} 
 (days)  & Healthy & (I) & (II)  & (I) \& (II) \\
\hline
100 & 54 & 12 & 1 & 0 \\
200 & 36 & 24 & 3 & 5 \\
300 & 13 & 35 & 1 & 17 \\
400 & 0  & 13 & 2 & 28 \\
500 & 0  & 3  & 1 & 35 \\
600 & 0  & 0  & 1 & 30 \\
700 & 0  & 0  & 1 & 28 \\
\hline
\end{tabular}
\end{minipage}

\end{table}

\subsection{Mice tumor toxicological data \label{sec:example2}}

In 1974, the National Center for Toxicological Research conducted a large-scale experiment to investigate the toxicological effects of 2-Acetylaminofluorene (2-AAF), a synthetic compound known to cause tumors following metabolic activation. A total of 24,000 female mice were randomly assigned to either a control group or one of seven treatment groups, each exposed to varying dose levels of 2-AAF. The chemical was applied to different parts of the animals’ bodies. A key focus of analysis, as presented in \cite{Lindsey1993}, was on the group exposed to the highest dosage—150 parts per million. Tumor development was monitored at three distinct time points: 12, 18, and 33 months.

A more detailed analysis was conducted on a subset of 990 mice from both the high-dose and control groups, with specific attention given to the presence or absence of bladder tumors. These data include the number of animals that died or were sacrificed, and whether or not they exhibited tumors. The outcomes were treated as two distinct failure events—death (I) and tumor development (II). 

\begin{table}[ht]
\small
  \setlength\tabcolsep{5pt}\def\arraystretch{1.2}
\centering
\caption{Summary of mice experiment data by dose level and time}
\vspace{9pt}
\begin{tabular}{|c|c| c|r r r r|}
\hline
&Dose & Time   & \multicolumn{4}{c|}{Mice status} \\ \cline{4-7} 
Group& (ppm) & (days)  & Healthy & (I) & (II)  & (I) \& (II) \\
\hline
1 & 0   & 12 & 23  & 0  & 3  & 3  \\
2 & 0   & 18 & 156 & 0  & 9  & 1  \\
3 & 0   & 33 & 134 & 1  & 49 & 8  \\
4 & 150 & 12 & 22  & 0  & 7  & 6  \\
5 & 150 & 18 & 73  & 35 & 4  & 12 \\
6 & 150 & 33 & 64  & 38 & 3  & 20 \\
\hline
\end{tabular}
\label{tab:mouse_data}
\end{table}

\section{Modeling the dependence among failure modes \label{sec:Copulas}}

In one-shot device testing with multiple failure modes, the classical competing risks model assumes that failures occur independently across modes, with each failure mechanism acting in isolation \cite{Balakrishnan2015a,Balakrishnan2015b,Balakrishnan2024}. While this simplifies analysis, the assumption is often unrealistic in practice, as failure in one mode may influence the likelihood of failure in another. This interdependence motivates the need for more flexible models that can capture such dependencies explicitly.

To account for the dependence structure between failure modes, we adopt a copula-based modeling approach. A copula links marginal cumulative distribution functions (CDFs) into a joint distribution, capturing the dependence structure separately from the marginal behavior. A two-dimensional copula $C(u,v)$, where $u,v \in [0,1]$, is defined via the Sklar's theorem:
$$
C(u, v) = P(U \le u, V \le v) = H(F^{-1}(u), G^{-1}(v)),
$$
with marginals $F$ and $G$ and joint CDF $H$. 

We consider two special cases of Archimedean copulas commonly used in reliability analysis: the Gumbel-Hougaard (GH) copula, which captures strong upper tail dependence and is suitable for modeling asymmetric dependencies; and the Frank copula, which is symmetric and can model both positive and negative dependence across the full range of the distribution, though it lacks tail dependence. Expressions for the most important parametric families of Archimedean copulas can be found in Table 4.1 in Nelsen \cite{Nelsen2006}.

\subsubsection*{Gumbel-Hougaard copula}

\begin{figure}[h]
\centering
\begin{tabular}{ll}
\includegraphics[scale=0.48, trim=0.25in 0 0 0, clip]{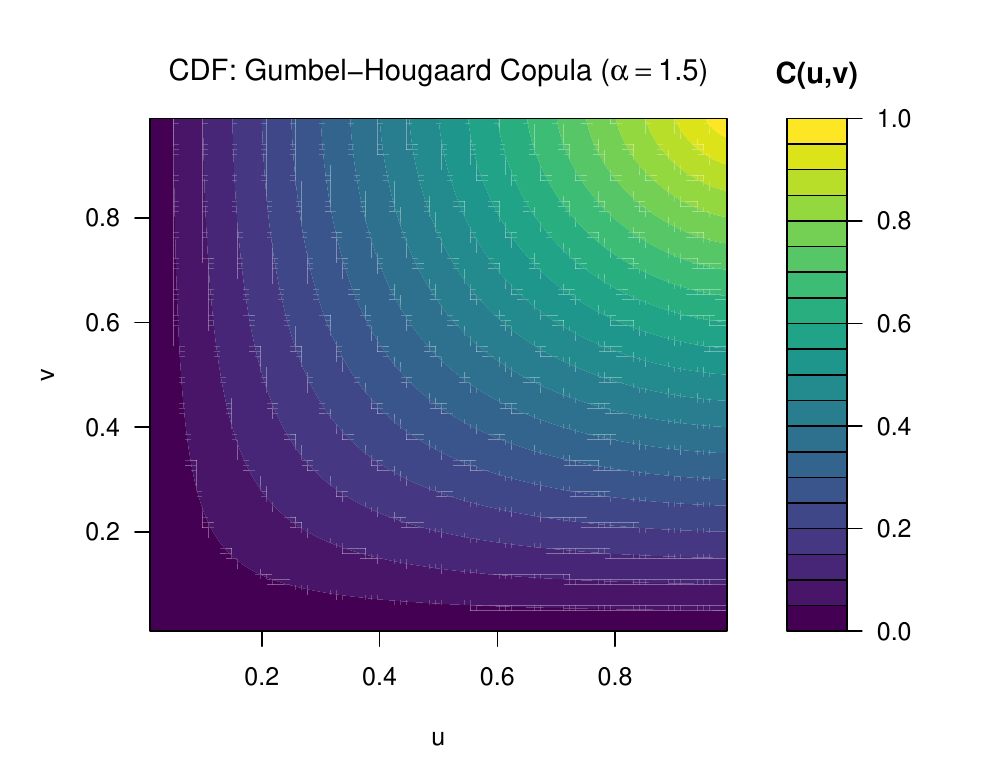}&
\includegraphics[scale=0.48, trim=0.25in 0 0 0, clip]{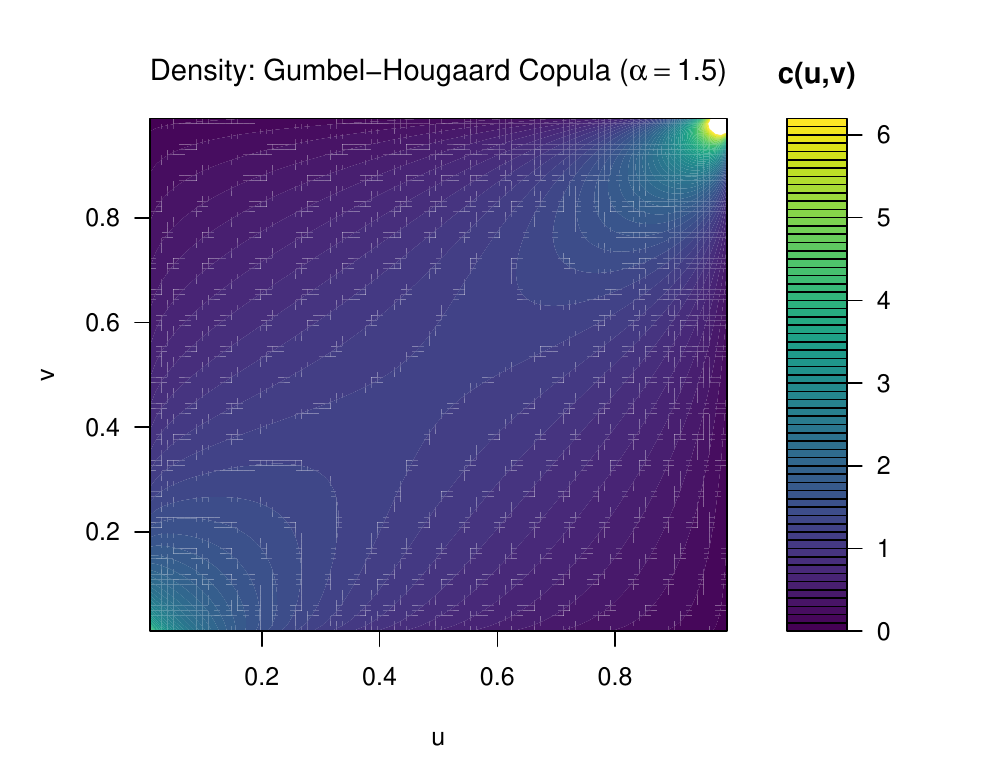}
\end{tabular}
\caption{CDF and density of GH copula for $\alpha=1.5$. \label{fig:GH}}
\end{figure}

The GH copula is suitable for modeling positive dependence, especially when extreme values (i.e., simultaneous late failures) are more likely. Its form is:
$$
C_{\alpha}(u, v) = \exp\left( -\left[(-\log u)^{\alpha} + (-\log v)^{\alpha}\right]^{1/\alpha} \right),
$$
with $0\leq u \leq 1$, $0\leq v \leq 1$, and the density is given by:
\begin{align*}
c_{\alpha}(u, v) =& C(u, v) \times \frac{ \left[ (-\log u)^\alpha + (-\log v)^\alpha \right]^{\frac{2}{\alpha} - 2} }{uv \cdot (\log u \log v)^{\alpha - 1} } \\
& \times \left[ (\alpha - 1)(-\log u)^\alpha (-\log v)^\alpha + (-\log u)^\alpha + (-\log v)^\alpha \right],
\end{align*}
where \( \alpha \ge 1 \) is the dependence parameter. When $ \alpha = 1$, the failure modes are independent. As \( \alpha \to \infty \), the dependence becomes stronger. In particular, the strength of association can be expressed via Kendall’s tau:
\begin{align}\label{eq:GH_tau}
\tau = 1 - \frac{1}{\alpha}.
\end{align}
As shown in the contour plots in Figure~\ref{fig:GH}, the high values near the top-right corner of the CDF (left) reflect strong upper tail dependence. The density plot (right) further emphasizes this behavior, with a peak near $u = v = 1$, illustrating the copula's ability to model simultaneous late failures.

\subsubsection*{Frank copula}

The Frank copula allows for both positive and negative dependence and is defined as:
$$
C_{\alpha}(u, v) = -\frac{1}{\alpha} \log\left(1 + \frac{(e^{-\alpha u} - 1)(e^{-\alpha v} - 1)}{e^{-\alpha} - 1} \right),
$$
with $0\leq u \leq 1$, $0\leq v \leq 1$, and the density is given by:
$$
c_{\alpha}(u, v) = \frac{e^{(1+u+v)\alpha}(e^{\alpha}-1)\alpha}{\left(e^{(u+v)\alpha} -e^{\alpha}(e^{u\alpha}+e^{v\alpha}-1)\right)^2} ,
$$
when $\alpha \in \mathbb{R} \setminus \{0\}$. When $\alpha = 0$, the copula reduces to the independence case. The Frank copula is useful for data without strong tail dependence and provides a symmetric dependence structure. Kendall’s tau is related to \( \alpha \) via:
\begin{align}\label{eq:tauFrank}
\tau = 1 + \frac{4}{\alpha} \left( \frac{1}{\alpha} \int_0^{\alpha} \frac{t}{e^t - 1} dt - 1 \right).
\end{align}

\begin{figure}[h]
\centering
\begin{tabular}{ll}
\includegraphics[scale=0.48, trim=0.25in 0 0 0, clip]{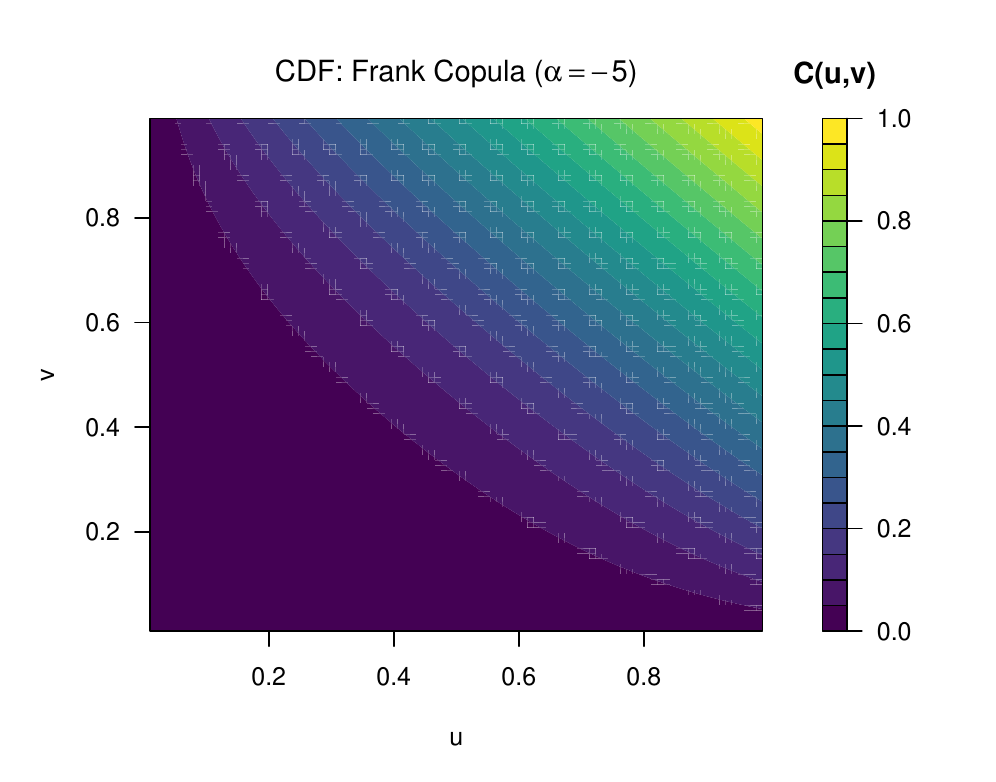}&
\includegraphics[scale=0.48, trim=0.25in 0 0 0, clip]{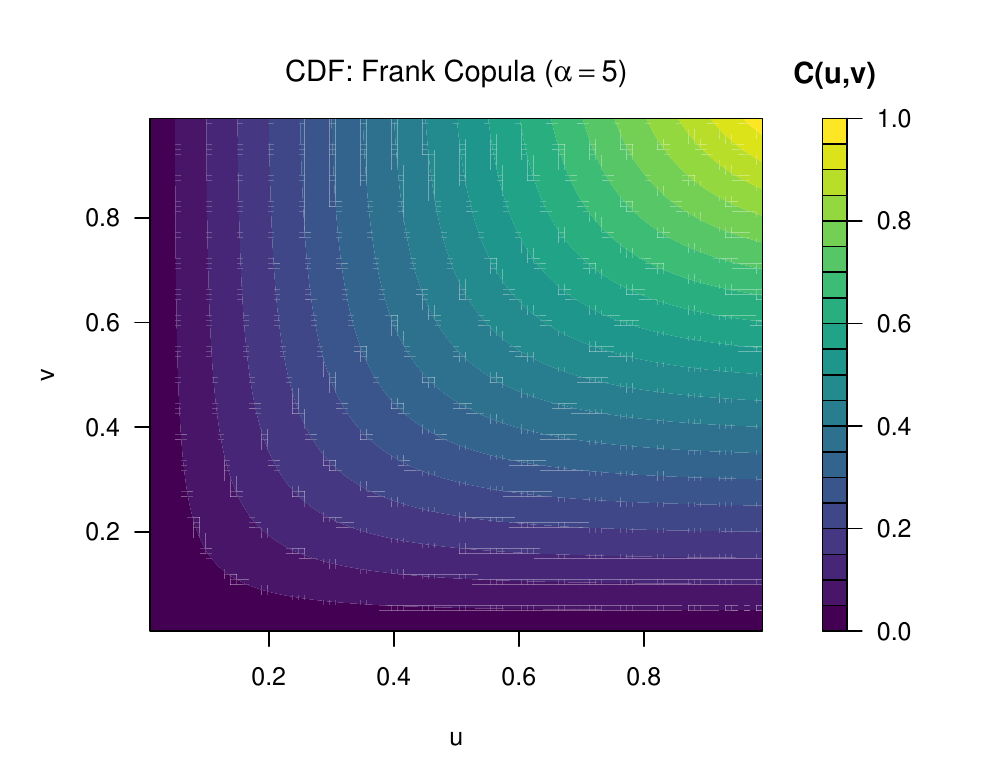}\\
\includegraphics[scale=0.48, trim=0.25in 0 0 0, clip]{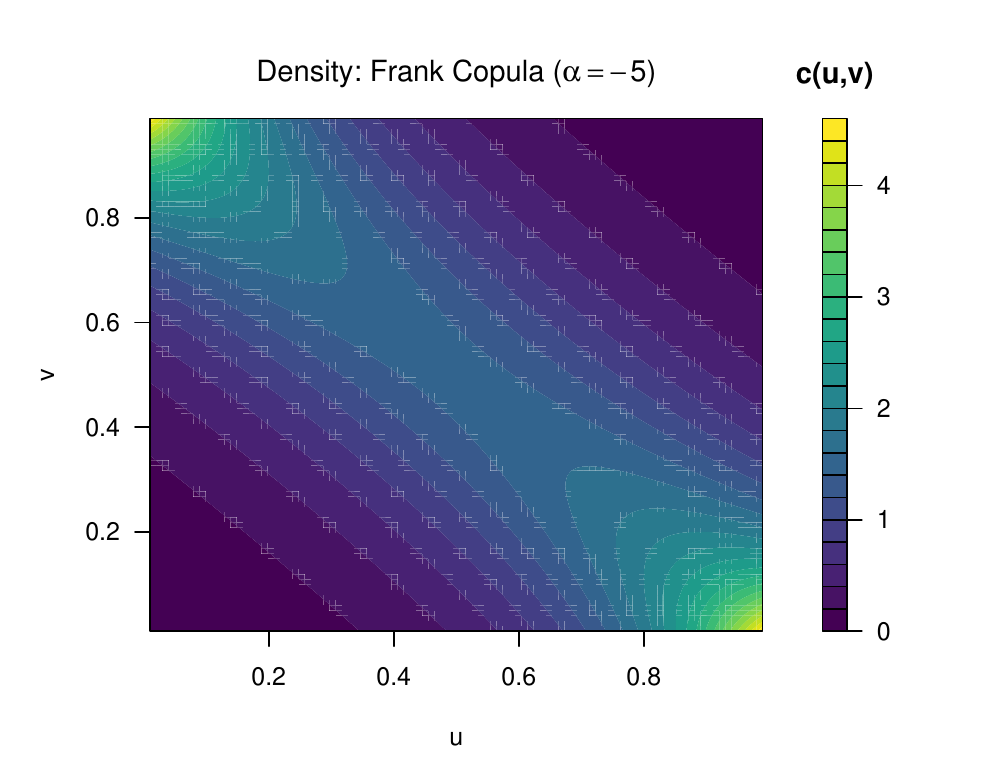}&
\includegraphics[scale=0.48, trim=0.25in 0 0 0, clip]{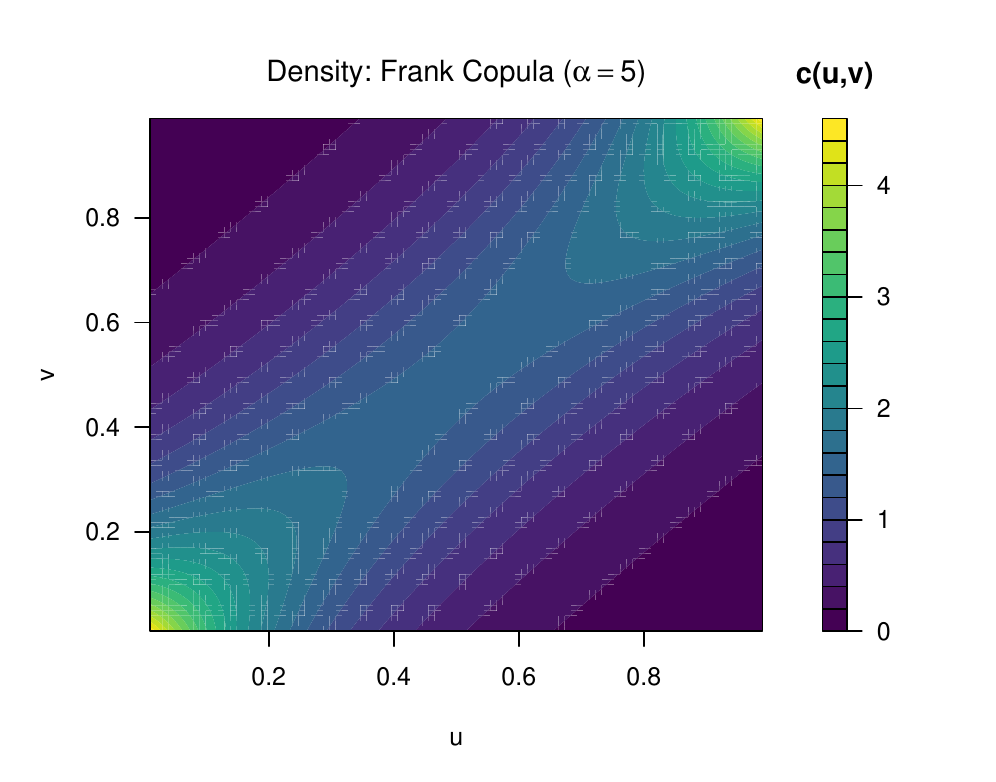}
\end{tabular}
\caption{CDF and density of Frank copula for $\alpha=-5$ (left) and $\alpha=5$ (right).\label{fig:frank}}
\end{figure}

The behavior of the CDF and density of the Frank copula for different values of $\alpha$ (i.e., $\alpha = \pm 5$) is illustrated in the contour plots in Figure~\ref{fig:frank}. The density plots (bottom row) illustrate the dependence structure of the Frank copula. In the positive dependence case ($\alpha=5$), the density is concentrated along the diagonal, indicating a tendency for both failure modes to occur together. In contrast, the negative dependence case ($\alpha=-5$) shows peaks in opposite corners, reflecting scenarios where one failure occurs early and the other late.

\subsection{Application to one-shot device testing \label{sec:application}}

Let $F_{ij,1} = F_1(IT_i, x_j)$ and $F_{ij,2} = F_2(IT_i, x_j)$ denote the marginal CDFs for failure modes 1 and 2, respectively, at inspection time $IT_i$ and stress level $x_j$. We express the joint distribution with a bivariate copula where the dependence parameter is allowed to vary with stress
\begin{align*}
\alpha_j=g(a_0+a_1x_j).
\end{align*}
Particularly,  for the GH copula $g(y)=1+e^y$,  guarantying $\alpha_j\geq 1$:

\begin{align*}
\alpha_j\equiv \alpha_j(\boldsymbol{\theta})=1+\exp(a_0+a_1x_j),
\end{align*}
while for the Frank copula,  $g(y)=y$, as $\alpha_j \in \mathbb{R}$:
\begin{align*}
\alpha_j\equiv  \alpha_j(\boldsymbol{\theta}) = a_0 + a_1 x_j,
\end{align*}
where $\boldsymbol{\theta}=(a_0,a_1)^T$ is the vector of parameters to be estimated. This formulation allows the dependence between the two failure modes to change across stress levels. We simplify the notation for clarity

$$
C_{\boldsymbol{\theta}}(i, j) \equiv  C_{\alpha_j(\boldsymbol{\theta})}(F_{ij,1}, F_{ij,2}) = P(T_1 \le IT_i, T_2 \le IT_i \mid x_j),
$$
where $T_1$ and $T_2$ denote the latent failure times associated with modes 1 and 2, respectively. Under independence, this reduces to $C_{\boldsymbol{\theta}}(i, j) = F_{ij,1} \times F_{ij,2}$. From this joint CDF, the probabilities of each failure category in the one-shot data are given by:
\begin{align*}
\pi_{ij,0}(\boldsymbol{\theta}) &= 1 - F_{ij,1} - F_{ij,2} + C_{\boldsymbol{\theta}}(i, j), \\
\pi_{ij,1}(\boldsymbol{\theta}) &= F_{ij,1} - C_{\boldsymbol{\theta}}(i, j), \\
\pi_{ij,2}(\boldsymbol{\theta}) &= F_{ij,2} - C_{\boldsymbol{\theta}}(i, j), \\
\pi_{ij,12}(\boldsymbol{\theta}) &= C_{\boldsymbol{\theta}}(i, j),
\end{align*}
where each $\pi_{ij,r}$ represents the probability of a specific failure pattern (no failure, mode 1 only, mode 2 only, or both).

\section{Estimation methods  \label{sec:Estimation}}

In this section, we discuss the methods for estimating the parameters associated with the chosen copula, $\boldsymbol{\theta}=(a_0,a_1)^T$. First, we present the classical maximum likelihood approach, followed by a divergence-based alternative that may offer a more robust estimation.

\vspace{-0.5cm}

\subsection{Classical maximum likelihood estimation approach}

The observed log-likelihood function in this case is given by
\begin{align*}
    \ell^\ast_{\text{obs}}(\boldsymbol{\theta}) = \sum_{i=1}^I \sum_{j=1}^J & \, n_{ij,0} \log(\pi_{ij,0}(\boldsymbol{\theta})) + n_{ij,1} \log(\pi_{ij,1}(\boldsymbol{\theta})) \notag \\
    & \quad + n_{ij,2} \log(\pi_{ij,2}(\boldsymbol{\theta})) + n_{ij,12} \log(\pi_{ij,12}(\boldsymbol{\theta})).
\end{align*}

\cite{Ling2021} proposed a two-step procedure for parameter estimation. In the first step, the marginal distributions are estimated using the observed proportion of failures, as follows:
\begin{align}\label{eq:Fij}
\widehat{F}_{ij,r} = \frac{N_{ij,r}}{K_{ij}},
\end{align}
and these estimates are then plugged into the log-likelihood to obtain a composite log-likelihood function:
\begin{align}
    \ell_{\text{obs}}(\boldsymbol{\theta}) = \sum_{i=1}^I \sum_{j=1}^J & \, n_{ij,0} \log\left(1 - \widehat{F}_{ij,1} - \widehat{F}_{ij,2} + C_{\boldsymbol{\theta}}(i,j)\right) \notag \\
    & \quad + n_{ij,1} \log\left(\widehat{F}_{ij,1} - C_{\boldsymbol{\theta}}(i,j)\right) \notag \\
    & \quad + n_{ij,2} \log\left(\widehat{F}_{ij,2} - C_{\boldsymbol{\theta}}(i,j)\right) + n_{ij,12} \log(C_{\boldsymbol{\theta}}(i,j)) \label{eq:COPULA_log_lik_composite}.
\end{align}
The  quasi-MLE (QMLE) of the parameter vector $\boldsymbol{\theta}$ associated with the chosen copula can then be estimated by maximizing (\ref{eq:COPULA_log_lik_composite}) with respect to $\boldsymbol{\theta}$, i.e.,
\begin{align}\label{eq:MLE}
\widehat{\boldsymbol{\theta}}_{\text{QMLE}} = \text{arg max}_{\boldsymbol{\theta}} \ \ell_{\text{obs}}(\boldsymbol{\theta}).
\end{align}

\subsection{A new divergence-based estimation approach}

Consider, for the $ij$-th testing condition, the empirical probability vector determined by the observed data:
\begin{align*}
\boldsymbol{p}_{ij} = \left( \frac{n_{ij,0}}{K_{ij}}, \frac{n_{ij,1}}{K_{ij}}, \frac{n_{ij,2}}{K_{ij}}, \frac{n_{ij,12}}{K_{ij}} \right)^T,
\end{align*}
and the theoretical probability vector determined by the model:
\begin{align}\label{eq:vector_ptheta}
\boldsymbol{\pi}_{ij}(\boldsymbol{\theta}) = \left( \pi_{ij,0}(\boldsymbol{\theta}), \pi_{ij,1}(\boldsymbol{\theta}), \pi_{ij,2}(\boldsymbol{\theta}), \pi_{ij,12}(\boldsymbol{\theta}) \right)^T.
\end{align}

The marginal distributions are estimated using the observed proportion of failures, as in (\ref{eq:Fij}), and an estimated value of (\ref{eq:vector_ptheta}) is given by
\begin{align*}
\widehat{\boldsymbol{\pi}}_{ij}(\boldsymbol{\theta}) &= \left( \widehat{\pi}_{ij,0}(\boldsymbol{\theta}), \widehat{\pi}_{ij,1}(\boldsymbol{\theta}), \widehat{\pi}_{ij,2}(\boldsymbol{\theta}), \widehat{\pi}_{ij,12}(\boldsymbol{\theta}) \right)^T \notag \\
&= \left( 1 - \widehat{F}_{ij,1} - \widehat{F}_{ij,2} + C_{\boldsymbol{\theta}}(i,j), \widehat{F}_{ij,1} - C_{\boldsymbol{\theta}}(i,j), \widehat{F}_{ij,2} - C_{\boldsymbol{\theta}}(i,j), C_{\boldsymbol{\theta}}(i,j) \right)^T. 
\end{align*}

\begin{definition}
The Kullback-Leibler (KL) divergence from $\widehat{\boldsymbol{\pi}}_{ij}(\boldsymbol{\theta})$ to $\boldsymbol{p}_{ij}$, $d_{\text{KL}}(\boldsymbol{p}_{ij}, \widehat{\boldsymbol{\pi}}_{ij}(\boldsymbol{\theta}))$, is defined as
\begin{align}
d_{\text{KL}}(\boldsymbol{p}_{ij}, \widehat{\boldsymbol{\pi}}_{ij}(\boldsymbol{\theta})) &= \frac{1}{K_{ij}} \left\{ n_{ij,0} \log\left( \frac{n_{ij,0}/K_{ij}}{\widehat{\pi}_{ij,0}(\boldsymbol{\theta})} \right) + n_{ij,1} \log\left( \frac{n_{ij,1}/K_{ij}}{\widehat{\pi}_{ij,1}(\boldsymbol{\theta})} \right) \right. \notag \\
& \quad \quad \quad \quad + n_{ij,2} \log\left( \frac{n_{ij,2}/K_{ij}}{\widehat{\pi}_{ij,2}(\boldsymbol{\theta})} \right) + n_{ij,12} \log\left( \frac{n_{ij,12}/K_{ij}}{\widehat{\pi}_{ij,12}(\boldsymbol{\theta})} \right) \Big\}, \label{eq:dk}
\end{align}
and the weighted KL divergence for all the units is defined as
\begin{align}\label{eq:DK}
D^{W}_{\text{KL}}(\boldsymbol{\theta}) = \sum_{i=1}^I \sum_{j=1}^J \frac{K_{ij}}{K} d_{\text{KL}}(\boldsymbol{p}_{ij}, \widehat{\boldsymbol{\pi}}_{ij}(\boldsymbol{\theta})).
\end{align}
\end{definition}

\begin{proposition}\label{Prop:MLE}
The QMLE of the parameter vector associated with the copula defined in (\ref{eq:MLE}) can alternatively be obtained by minimizing the KL divergence for all units, as defined in (\ref{eq:DK}), i.e.,
\begin{align*}
\widehat{\boldsymbol{\theta}}_{\text{QMLE}} = \text{arg min}_{\boldsymbol{\theta}} \, D^{W}_{\text{KL}}(\boldsymbol{\theta}).
\end{align*}
\end{proposition}
\begin{proof}
 Proposition \ref{Prop:MLE} is easily proved by considering the form of the KL divergence from $\widehat{\boldsymbol{\pi}}_{ij}(\boldsymbol{\theta})$ to $\boldsymbol{p}_{ij}$, given in (\ref{eq:dk}), which can be expressed as
\begin{align*}
d_{\text{KL}}(\boldsymbol{p}_{ij}, \widehat{\boldsymbol{\pi}}_{ij}(\boldsymbol{\theta})) &= \frac{1}{K_{ij}} \left\{ n_{ij,0} \log\left( \frac{n_{ij,0}/K_{ij}}{\widehat{\pi}_{ij,0}(\boldsymbol{\theta})} \right) + n_{ij,1} \log\left( \frac{n_{ij,1}/K_{ij}}{\widehat{\pi}_{ij,1}(\boldsymbol{\theta})} \right) \right.\\
& \quad \quad \quad \quad + n_{ij,2} \log\left( \frac{n_{ij,2}/K_{ij}}{\widehat{\pi}_{ij,2}(\boldsymbol{\theta})} \right) + n_{ij,12} \log\left( \frac{n_{ij,12}/K_{ij}}{\widehat{\pi}_{ij,12}(\boldsymbol{\theta})} \right) \Big\}, \\
&= \frac{1}{K_{ij}} \left\{ n_{ij,0} \log\left( n_{ij,0}/K_{ij} \right) + n_{ij,1} \log\left( n_{ij,1}/K_{ij} \right) \right.  \\
& \quad \quad \quad \quad + n_{ij,2} \log\left( n_{ij,2}/K_{ij} \right) + n_{ij,12} \log\left( n_{ij,12}/K_{ij} \right) \}, \\
&\quad-\frac{1}{K_{ij}} \left\{ n_{ij,0} \log\left( \widehat{\pi}_{ij,0}(\boldsymbol{\theta}) \right) + n_{ij,1} \log\left(\widehat{\pi}_{ij,1}(\boldsymbol{\theta}) \right) \right.\\
 &\quad \quad \quad \quad + n_{ij,2} \log\left( \widehat{\pi}_{ij,2}(\boldsymbol{\theta}) \right) + n_{ij,12} \log\left( \widehat{\pi}_{ij,12}(\boldsymbol{\theta})\right) \}
\end{align*}
This can be rewritten as:
\begin{align*}
d_{\text{KL}}(\boldsymbol{p}_{ij}, \widehat{\boldsymbol{\pi}}_{ij}(\boldsymbol{\theta})) &= C - \frac{1}{K_{ij}} \Big\{ n_{ij,0} \log(\widehat{\pi}_{ij,0}(\boldsymbol{\theta})) + n_{ij,1} \log(\widehat{\pi}_{ij,1}(\boldsymbol{\theta})) \\
& \quad + n_{ij,2} \log(\widehat{\pi}_{ij,2}(\boldsymbol{\theta})) + n_{ij,12} \log(\widehat{\pi}_{ij,12}(\boldsymbol{\theta})) \Big\},
\end{align*}
where 
\begin{align*}
  C&= \frac{1}{K_{ij}} \left\{ n_{ij,0} \log\left( n_{ij,0}/K_{ij} \right) + n_{ij,1} \log\left( n_{ij,1}/K_{ij} \right) \right.  \\
& \quad \quad \quad \quad + n_{ij,2} \log\left( n_{ij,2}/K_{ij} \right) + n_{ij,12} \log\left( n_{ij,12}/K_{ij} \right) \}  
\end{align*}
is a constant that does not depend on $\boldsymbol{\theta}$.    Therefore, the maximization of (\ref{eq:COPULA_log_lik_composite}) is equivalent to the minimization of the weighted KL divergence defined in (\ref{eq:DK}), as required.
\end{proof}

Once we have established that the QMLE can be obtained by minimizing the Kullback-Leibler (KL) divergence between the empirical and theoretical probability vectors associated with the model, it is natural to consider alternative divergence measures for obtaining different estimators. While QMLE is known for its efficiency, it is also criticized for its lack of robustness in the presence of outliers. Therefore, it is preferable to explore alternative divergences that might lead to more robust estimators. In this regard, density power divergence (DPD) \cite{Basu1998} has been extensively studied in the literature, showing excellent results in terms of robustness with minimal loss in efficiency. In this work, we propose DPD-based estimators for estimating the parameters associated with copulas in one-shot device testing data.

\begin{definition}
Given the tuning parameter $\beta \geq 0$, the weighted DPD measure for all units is defined as
\begin{align*}
D^{W}_{\beta}(\boldsymbol{\theta}) = \sum_{i=1}^I \sum_{j=1}^J \frac{K_{ij}}{K} d_{\beta}(\boldsymbol{p}_{ij}, \widehat{\boldsymbol{\pi}}_{ij}(\boldsymbol{\theta})),
\end{align*}
where, for $\beta = 0$,
\[
d_{\beta=0}(\boldsymbol{p}_{ij}, \widehat{\boldsymbol{\pi}}_{ij}(\boldsymbol{\theta})) = d_{\text{KL}}(\boldsymbol{p}_{ij}, \widehat{\boldsymbol{\pi}}_{ij}(\boldsymbol{\theta})),
\]
and for $\beta > 0$,
\begin{align*}
&d_{\beta}(\boldsymbol{p}_{ij}, \widehat{\boldsymbol{\pi}}_{ij}(\boldsymbol{\theta})) = \ \widehat{\pi}_{ij,0}^{1+\beta}(\boldsymbol{\theta}) + \widehat{\pi}_{ij,1}^{1+\beta}(\boldsymbol{\theta}) + \widehat{\pi}_{ij,2}^{1+\beta}(\boldsymbol{\theta}) + \widehat{\pi}_{ij,12}^{1+\beta}(\boldsymbol{\theta}) \\
& \quad \quad - \frac{1+\beta}{\beta} \left\{ \frac{n_{ij,0}}{K_{ij}} \widehat{\pi}_{ij,0}^{\beta}(\boldsymbol{\theta}) + \frac{n_{ij,1}}{K_{ij}} \widehat{\pi}_{ij,1}^{\beta}(\boldsymbol{\theta}) + \frac{n_{ij,2}}{K_{ij}} \widehat{\pi}_{ij,2}^{\beta}(\boldsymbol{\theta}) + \frac{n_{ij,12}}{K_{ij}} \widehat{\pi}_{ij,12}^{\beta}(\boldsymbol{\theta}) \right\}.
\end{align*}
\end{definition}

\vspace{12pt}

Thus, the DPD depends on the tuning parameter $\beta \geq 0$, resulting in a family of divergences that includes the KL divergence as a special case when $\beta = 0$. Therefore, we define the family of quasi minimum DPD estimators (QMDPDE) for the parameter associated with the chosen copula as follows:
\begin{align*}
\widehat{\boldsymbol{\theta}}_{\beta} = \text{arg min}_{\boldsymbol{\theta}} \, D^{W}_{\beta}(\boldsymbol{\theta}).
\end{align*}

In Section  \ref{sec:numerical}, we will show through an extensive simulation study that using QMDPDEs with $\beta > 0$ may lead to more robust estimation than that obtained with the QMLE.

Although the density power divergence is continuous at $\beta = 0$ and coincides with the KL divergence in the limit, the behavior for very small positive values of $\beta$ may be numerically unstable in practice. This is due to the presence of terms involving $1/\beta$ in the divergence formulation, which can lead to ill-conditioned optimization problems when $\beta$ is close to zero.  Our numerical experiments indicate that extremely small values (e.g., $\beta = 0.0001$) may actually result in less stable estimation than moderate values, and do not necessarily approximate the QMLE well in practice. In contrast, small but not negligible values of $\beta$ (e.g., $\beta = 0.1$) provide a much more stable behavior while still remaining close to the QMLE. Therefore, in practice, it is preferable to either use $\beta = 0$ (QMLE) or moderate positive values of $\beta$, avoiding extremely small values.

\section{Implementation and numerical study \label{sec:numerical}}

\subsection{Optimization algorithm and initial  parameters}
Since the parameter estimates cannot be derived in a closed-form expression, optimization techniques, such as \texttt{optim()} or \texttt{nlm()} in R, are employed to estimate these parameters.

 To compute the integral in (\ref{eq:tauFrank}) for the Frank Copula, we can use the function \texttt{integrate()} in \texttt{R}. To avoid computing this integral, the approximation
\begin{align}\label{eq:approx_Frank}
\tau \approx \frac{\alpha_F}{9}
\end{align}
 performs well when $-3 \leq\alpha_F \leq 3$, or equivalently, $-0.307 \leq \tau \leq 0.307$.   More details are given in \cite{Ling2021}.\\

For selecting the initial values, we may use a first approximation based on Kendall's tau (as defined by Nelson \cite{Nelson2009}), which, for each $j=1,\dots,J$ can be expressed as:
$$
\widehat{\tau}_j^{(0)}=\frac{C_j-D_j}{C_j+D_j},
$$
where $C_j$ is the number of concordant pairs, and $D_j$ is the number of discordant pairs at stress level $x_j$. In the context of one-shot device testing, these are given as
$$
C_j= \sum_{i=1}^I n_{ij,0}n_{ij,12}, \ \quad D_j=\sum_{i=1}^I n_{ij,1}n_{ij,2}. 
$$
The relationships  between the Archimedean copula parameters and Kendall’s tau in (\ref{eq:GH_tau}) and (\ref{eq:approx_Frank}) can be utilized to derive initial estimates for the copula parameters $\widehat{\alpha}_j^{(0)}$. Then, for determining the starting values of $(a_0,a_1)$, the least-squares approach is employed, defined as:
$$
\boldsymbol{\theta}^{(0)}=\left(\boldsymbol{X}^T\boldsymbol{X} \right)^{-1}\boldsymbol{X}^T\boldsymbol{\Lambda}^{(0)},
$$
where $\boldsymbol{\theta}^{(0)}=(\widehat{a}_0^{(0)},\widehat{a}_1^{(0)})^T$, and
$$
\boldsymbol{X}  = \begin{pmatrix}
1 & x_1 \\
1 & x_2 \\
\vdots & \vdots \\
1 & x_J
\end{pmatrix}_{J \times 2}, \quad
\boldsymbol{\Lambda}^{(0)}  = \begin{pmatrix}
\lambda_1^{(0)} \\
\lambda_2^{(0)} \\
\vdots \\
\lambda_J^{(0)}
\end{pmatrix}_{J \times 1},
$$
where $\lambda_j^{(0)}=g^{-1}(\widehat{\alpha}_j^{(0)})$, with $j=1,\dots,J$, with $g(\cdot)$ defined as in Section \ref{sec:application} for the different copulas.

\subsection{Monte Carlo simulation scheme}

Here, we develop an extensive simulation study to evaluate the performance of the proposed estimation methods under both GH and Frank copulas. In both scenarios, we aim to emulate realistically motivated situations—one inspired by automotive airbag testing and the other by toxicological sacrifice experiments—allowing us to assess estimator performance under conditions that resemble real-world applications.

\subsubsection{First simulation scenario: emulating automotive airbag testing}

To provide a coherent and practically motivated setting for evaluating the proposed estimators, we reinterpret the parameter structure used in \cite{Balakrishnan2021} (limited to MLE) in the context of accelerated testing for one–shot devices, such as automotive airbags. Although the cited paper introduces these parameter configurations purely for methodological illustration, here we adopt them as a stylized representation loosely inspired by how thermal stress may influence the occurrence of multiple failure modes in airbags.

\begin{figure}
    \centering
    \includegraphics[scale=0.2]{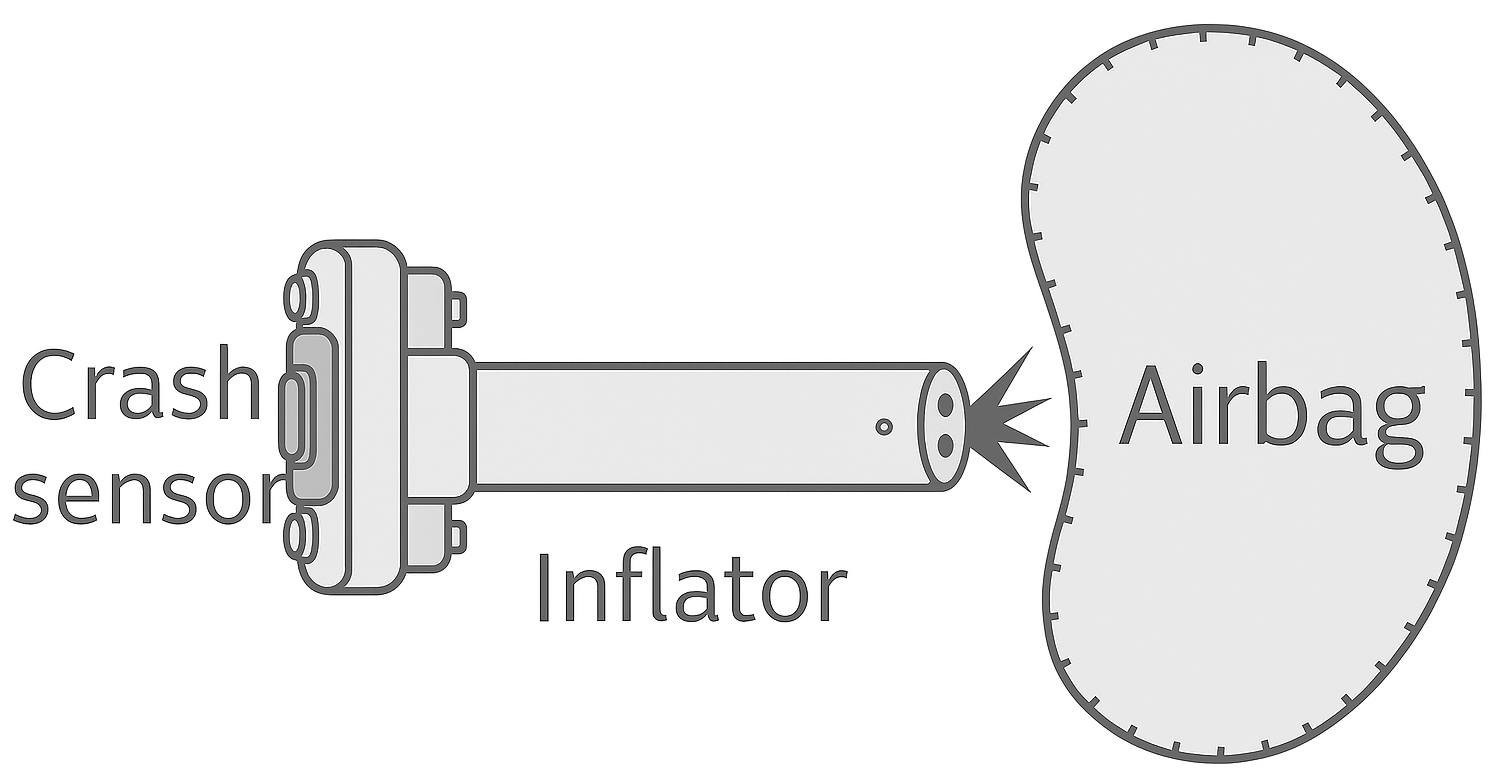}
    \caption{Simplified schematic representation (not to scale) of an airbag system showing crash sensor, inflator, and airbag.}
    \label{fig:airbag}
\end{figure}

Three nominal temperature levels, $30^\circ$C, $40^\circ$C, and $50^\circ$C, are taken as the stress conditions under which devices are inspected. As temperature varies, both the crash sensor electronics and the inflator housing may become more susceptible to degradation, leading to two observable failure modes:  sensor failure (Mode 1) and inflator rupture (Mode 2). See Figure \ref{fig:airbag} for a simplified schematic illustration.

For each stress–inspection configuration, a total of $K^*$ one–shot items are tested, with $K^* \in \{50, 100, 200\}$ representing low, moderate, and high sample sizes. The lifetime distribution of failure modes is generated from the following parametric models:

\begin{enumerate}
\item[i)] Weibull distribution with common scale parameter $\beta_j=\exp(s_0+s_1x_j)$ and shape parameter for each failure mode $\eta_{j,g}=\exp(r_{0,g}+r_1x_j)$, with $(s_0,s_1,r_{0,1},r_{0,2},r_1)=(3.5,-0.02,2,2.1,-0.03)$,
\item[ii)] Gamma distribution with common scale parameter $\beta_j=\exp(s_0+s_1x_j)$ and shape parameter for each failure mode $\eta_{j,g}=\exp(r_{0,g}+r_1x_j)$, with $(s_0,s_1,r_{0,1},r_{0,2},r_1)=(-0.3,0.04,3.6,3.8,-0.06)$.
\end{enumerate}
These parameter choices produce lifetime distributions that react to changes in stress in a controlled and smooth way, without assuming any specific physical law linking temperature and failure probability, and ensuring distinct yet comparable degradation patterns across both distributional families.

Regarding dependence among failure modes, the parameters of interest $\boldsymbol{\theta}=(a_0,a_1)^T$ under a) GH copula; b) Frank copula with positive dependence; and c) Frank copula with negative dependence, are taken to be $(-2,0.02)$, $(1,0.02)$ and $(-1,-0.02)$, respectively.

In this context, positive dependence is used to represent situations where the occurrence of one failure mechanism may increase the likelihood of the other being observed in the same unit. Although the crash sensor electronics and the inflator housing rely on different physical mechanisms, system–level interactions (e.g., shared circuitry, timing faults, or manufacturing variability) can make their failures statistically correlated. The negative–dependence configuration, while less common, may arise in cases where the manifestation of one failure mode effectively precludes the appearance of the other. Both forms are therefore included to assess estimator performance across a broad but conceptually plausible spectrum of dependence behaviors.

To evaluate robustness, we introduce contamination at the most extreme testing condition ($x=50$, $IT=20$). For each of the 1000 Monte Carlo datasets generated under the clean model, the counts associated with this final stress level are modified according to
\begin{align}\label{eq:contaminationMC}
 (\tilde{n}_0, \tilde{n}_1, \tilde{n}_2, \tilde{n}_{12}) = (n_0, n_1, n_2 + n_{12}, 0).   
\end{align}
effectively eliminating the simultaneous failures observed at this condition and redistributing them as Mode 2 failures. This corresponds to contaminating approximately one–twelfth of the observations and intentionally distorts the inferred dependence. Such contamination mimics misclassification or diagnostic errors that tend to occur under the harshest testing conditions.

We obtain the QMDPDE of the model parameters under different tuning parameters $\beta\geq 0$, noting that for $\beta = 0$, we have the QMLE. The copula parameter $\alpha_0$ and the dependence parameter $\tau_0$ are evaluated under normal operating conditions, using $x_0=25$ as a nominal reference stress level. By computing the mean over 1000 samples, the estimates are presented in Tables \ref{table:GH}, \ref{table:FRANK_P}, and \ref{table:FRANK_N} for the GH copula, the Frank copula with positive correlation, and the Frank copula with negative correlation, respectively.

For each copula and for both lifetime distributions, it is shown that, under a non-contaminated setup, there is no clear difference between the chosen tuning parameters $\beta$. On the other hand, when a contaminated scenario is considered, the QMDPDE shows more robust behavior as $\beta$ increases, while the QMLE exhibits the worst performance. Therefore, this justifies the use of QMDPDE as an alternative to QMLE due to its robustness, without significant loss in efficiency.

\begin{table}[p!!]\setlength\tabcolsep{4.2pt}\renewcommand{\arraystretch}{1.2} 
\small
\centering
\caption{Estimated parameters for one-shot device models using the GH copula, under varying sample sizes ($K^* = 50, 100, 200$) and lifetime distributions (Weibull and Gamma). Results are shown for both non-contaminated and contaminated data, with different values of the tuning parameter $\beta$ in the QMDPDE. Estimates are compared with the QMLE ($\beta=0$). \label{table:GH}}
\vspace{12pt}
\begin{tabular}{|r|r|rrrr|rrrr|}
  \hline
 \multicolumn{2}{|c|}{Weibull lifetimes}& \multicolumn{4}{c|}{Non-contaminated data} &  \multicolumn{4}{c|}{Contaminated data} \\ 
  \hline
    & & & \multicolumn{3}{c|}{$\beta$} & & \multicolumn{3}{c|}{$\beta$}\\ \cline{4-6} \cline{8-10}
  Parameter & True value &QMLE & 0.2 & 0.4 & 0.6 & QMLE & 0.2 & 0.4 & 0.6  \\  \hline
  $K^*=50$ & &  &  &  &  &  &  &  &  \\
  $a_0$ & -2.000 & -2.359 & -2.475 & -2.523 & -2.525 & -0.711 & -1.435 & -1.904 & -2.149 \\ 
  $a_1$ & 0.020 & -0.009 & -0.007 & -0.003 & -0.002 & -0.059 & -0.037 & -0.013 & 0.001 \\ 
  $\alpha_0$ & 1.223 & 2.507 & 1.373 & 1.509 & 1.271 & 3.475 & 1.318 & 1.345 & 1.276 \\ 
  $\tau_0$ & 0.182 & 0.197 & 0.197 & 0.198 & 0.198 & 0.241 & 0.213 & 0.203 & 0.200 \\ \hline
  $K^*=100$ & &  &  &  &  &  &  &  &  \\
  $a_0$ & -2.000 & -2.080 & -2.061 & -2.065 & -2.062 & -0.817 & -1.416 & -1.746 & -1.871 \\ 
  $a_1$ & 0.020 & 0.013 & 0.011 & 0.011 & 0.011 & -0.024 & -0.002 & 0.008 & 0.012 \\ 
  $\alpha_0$ & 1.223 & 1.245 & 1.248 & 1.247 & 1.248 & 1.327 & 1.281 & 1.260 & 1.252 \\ 
  $\tau_0$ & 0.182 & 0.189 & 0.191 & 0.191 & 0.191 & 0.237 & 0.211 & 0.199 & 0.194 \\ \hline
  $K^*=200$ & &  &  &  &  &  &  &  &  \\
  $a_0$ & -2.000 & -2.039 & -2.013 & -2.017 & -2.020 & -0.837 & -1.452 & -1.784 & -1.914 \\ 
  $a_1$ & 0.020 & 0.020 & 0.020 & 0.020 & 0.020 & -0.015 & 0.004 & 0.013 & 0.017 \\ 
  $\alpha_0$ & 1.223 & 1.231 & 1.235 & 1.235 & 1.234 & 1.315 & 1.273 & 1.250 & 1.241 \\ 
  $\tau_0$ & 0.182 & 0.185 & 0.187 & 0.187 & 0.187 & 0.235 & 0.210 & 0.196 & 0.191  \\ \hline \hline
 \multicolumn{2}{|c|}{Gamma lifetimes}& \multicolumn{4}{c|}{Non-contaminated data} &  \multicolumn{4}{c|}{Contaminated data} \\ 
  \hline
    & & & \multicolumn{3}{c|}{$\beta$} & & \multicolumn{3}{c|}{$\beta$}\\ \cline{4-6} \cline{8-10}
  Parameter & True value &QMLE & 0.2 & 0.4 & 0.6 & QMLE & 0.2 & 0.4 & 0.6  \\  \hline
  $K^*=50$ & &  &  &  &  &  &  &  &  \\
  $a_0$ & -2.000 & -2.543 & -2.467 & -2.556 & -2.627 & 0.537 & -0.095 & -0.744 & -1.180 \\ 
  $a_1$ & 0.020 & -0.025 & -0.027 & -0.015 & -0.010 & -0.139 & -0.112 & -0.068 & -0.044 \\ 
  $\alpha_0$ & 1.223 & 1.263 & 1.264 & 1.266 & 1.267 & 2.671 & 2.049 & 1.366 & 1.535 \\ 
  $\tau_0$ & 0.182 & 0.192 & 0.193 & 0.194 & 0.195 & 0.244 & 0.212 & 0.208 & 0.206 \\ \hline
  $K^*=100$ & &  &  &  &  &  &  &  &  \\
  $a_0$ & -2.000 & -2.091 & -2.079 & -2.064 & -2.062 & 0.189 & -0.264 & -0.805 & -1.134 \\ 
  $a_1$ & 0.020 & 0.018 & 0.015 & 0.018 & 0.019 & -0.071 & -0.066 & -0.043 & -0.028 \\ 
  $\alpha_0$ & 1.223 & 1.241 & 1.243 & 1.245 & 1.245 & 1.364 & 1.311 & 1.286 & 1.274 \\ 
  $\tau_0$ & 0.182 & 0.188 & 0.188 & 0.190 & 0.190 & 0.254 & 0.225 & 0.211 & 0.204 \\ \hline
  $K^*=200$ & &  &  &  &  &  &  &  &  \\
  $a_0$ & -2.000 & -2.045 & -2.033 & -2.041 & -2.045 & -0.063 & -0.654 & -1.137 & -1.423 \\ 
  $a_1$ & 0.020 & 0.021 & 0.019 & 0.020 & 0.021 & -0.044 & -0.028 & -0.013 & -0.003 \\ 
  $\alpha_0$ & 1.223 & 1.233 & 1.235 & 1.234 & 1.234 & 1.366 & 1.318 & 1.284 & 1.267 \\ 
  $\tau_0$ & 0.182 & 0.185 & 0.186 & 0.186 & 0.186 & 0.262 & 0.235 & 0.216 & 0.205 \\ 
   \hline
\end{tabular}
\end{table}

\begin{table}[p!!]\setlength\tabcolsep{4.8pt}\renewcommand{\arraystretch}{1.2} 
\small
\centering
\caption{Estimated parameters for one-shot device models using the Frank copula with positive dependence, under varying sample sizes ($K^* = 50, 100, 200$) and lifetime distributions (Weibull and Gamma). Results are shown for both non-contaminated and contaminated data, with different values of the tuning parameter $\beta$ in the QMDPDE. Estimates are compared with the QMLE ($\beta=0$). \label{table:FRANK_P}}
\vspace{12pt}
\begin{tabular}{|r|r|rrrr|rrrr|}
  \hline
 \multicolumn{2}{|c|}{Weibull lifetimes}& \multicolumn{4}{c|}{Non-contaminated data} &  \multicolumn{4}{c|}{Contaminated data} \\ 
  \hline
    & & & \multicolumn{3}{c|}{$\beta$} & & \multicolumn{3}{c|}{$\beta$}\\ \cline{4-6} \cline{8-10}
  Parameter & True value &QMLE & 0.2 & 0.4 & 0.6 & QMLE & 0.2 & 0.4 & 0.6  \\
  \hline
  $K^*=50$ & &  &  &  &  &  &  &  &  \\
  $a_0$ & 1.000 & 1.108 & 1.128 & 1.134 & 1.145 & 3.275 & 2.350 & 1.806 & 1.546 \\ 
  $a_1$ & 0.020 & 0.019 & 0.019 & 0.019 & 0.018 & -0.045 & -0.017 & -0.001 & 0.006 \\ 
  $\alpha_0$ & 1.500 & 1.591 & 1.599 & 1.601 & 1.605 & 2.150 & 1.913 & 1.773 & 1.708 \\ 
  $\tau_0$ & 0.163 & 0.167 & 0.168 & 0.168 & 0.168 & 0.221 & 0.199 & 0.185 & 0.178 \\ \hline
  $K^*=100$ & &  &  &  &  &  &  &  &  \\
  $a_0$ & 1.000 & 0.977 & 0.987 & 0.977 & 0.983 & 3.108 & 2.124 & 1.562 & 1.300 \\ 
  $a_1$ & 0.020 & 0.022 & 0.021 & 0.022 & 0.021 & -0.042 & -0.013 & 0.004 & 0.012 \\ 
  $\alpha_0$ & 1.500 & 1.518 & 1.522 & 1.518 & 1.521 & 2.064 & 1.812 & 1.667 & 1.601 \\ 
  $\tau_0$ & 0.163 & 0.162 & 0.163 & 0.162 & 0.162 & 0.217 & 0.192 & 0.177 & 0.171 \\ \hline
  $K^*=200$ & &  &  &  &  &  &  &  &  \\
  $a_0$ & 1.000 & 0.989 & 1.008 & 0.996 & 0.993 & 3.139 & 2.141 & 1.570 & 1.303 \\ 
  $a_1$ & 0.020 & 0.020 & 0.020 & 0.020 & 0.020 & -0.044 & -0.014 & 0.003 & 0.011 \\ 
  $\alpha_0$ & 1.500 & 1.498 & 1.506 & 1.501 & 1.499 & 2.049 & 1.792 & 1.645 & 1.578 \\ 
  $\tau_0$ & 0.163 & 0.162 & 0.162 & 0.162 & 0.162 & 0.217 & 0.191 & 0.177 & 0.170   \\ \hline \hline
 \multicolumn{2}{|c|}{Gamma lifetimes}& \multicolumn{4}{c|}{Non-contaminated data} &  \multicolumn{4}{c|}{Contaminated data} \\ 
  \hline
    & & & \multicolumn{3}{c|}{$\beta$} & & \multicolumn{3}{c|}{$\beta$}\\ \cline{4-6} \cline{8-10}
  Parameter & True value &QMLE & 0.2 & 0.4 & 0.6 & QMLE & 0.2 & 0.4 & 0.6  \\  \hline
  $K^*=50$ & &  &  &  &  &  &  &  &  \\
  $a_0$ & 1.000 & 1.070 & 1.067 & 1.051 & 1.040 & 4.515 & 3.673 & 3.004 & 2.554 \\ 
  $a_1$ & 0.020 & 0.020 & 0.020 & 0.020 & 0.021 & -0.082 & -0.057 & -0.037 & -0.024 \\ 
  $\alpha_0$ & 1.500 & 1.568 & 1.566 & 1.559 & 1.554 & 2.476 & 2.254 & 2.075 & 1.955 \\ 
  $\tau_0$ & 0.163 & 0.164 & 0.164 & 0.163 & 0.162 & 0.252 & 0.230 & 0.213 & 0.201 \\\hline
  $K^*=100$ & &  &  &  &  &  &  &  &  \\ 
  $a_0$ & 1.000 & 1.060 & 1.074 & 1.064 & 1.053 & 4.456 & 3.598 & 2.902 & 2.429 \\ 
  $a_1$ & 0.020 & 0.019 & 0.019 & 0.019 & 0.019 & -0.081 & -0.056 & -0.035 & -0.021 \\ 
  $\alpha_0$ & 1.500 & 1.542 & 1.548 & 1.544 & 1.540 & 2.436 & 2.210 & 2.027 & 1.903 \\ 
  $\tau_0$ & 0.163 & 0.165 & 0.165 & 0.165 & 0.164 & 0.252 & 0.230 & 0.212 & 0.200 \\ \hline
  $K^*=200$ & &  &  &  &  &  &  &  &  \\
  $a_0$ & 1.000 & 1.037 & 1.043 & 1.028 & 1.022 & 4.412 & 3.545 & 2.841 & 2.341 \\ 
  $a_1$ & 0.020 & 0.020 & 0.020 & 0.020 & 0.020 & -0.080 & -0.054 & -0.034 & -0.019 \\ 
  $\alpha_0$ & 1.500 & 1.529 & 1.531 & 1.525 & 1.523 & 2.415 & 2.185 & 2.001 & 1.868 \\ 
  $\tau_0$ & 0.163 & 0.165 & 0.165 & 0.164 & 0.164 & 0.252 & 0.230 & 0.212 & 0.199 \\ 
   \hline
\end{tabular}
\end{table}

\begin{table}[p!!]\setlength\tabcolsep{4.5pt}\renewcommand{\arraystretch}{1.2} 
\small
\centering
\caption{Estimated parameters for one-shot device models using the Frank copula with negative dependence, under varying sample sizes ($K^* = 50, 100, 200$) and lifetime distributions (Weibull and Gamma). Results are shown for both non-contaminated and contaminated data, with different values of the tuning parameter $\beta$ in the QMDPDE. Estimates are compared with the QMLE ($\beta=0$).\label{table:FRANK_N}}
\vspace{12pt}
\begin{tabular}{|r|r|rrrr|rrrr|}
  \hline
 \multicolumn{2}{|c|}{Weibull lifetimes}& \multicolumn{4}{c|}{Non-contaminated data} &  \multicolumn{4}{c|}{Contaminated data} \\ 
  \hline
    & & & \multicolumn{3}{c|}{$\beta$} & & \multicolumn{3}{c|}{$\beta$}\\ \cline{4-6} \cline{8-10}
  Parameter & True value &QMLE & 0.2 & 0.4 & 0.6 & QMLE & 0.2 & 0.4 & 0.6  \\  \hline
  $K^*=50$ & &  &  &  &  &  &  &  &  \\
  $a_0$ & -1.000 & -1.051 & -1.087 & -1.101 & -1.116 & 0.731 & 0.388 & 0.111 & -0.112 \\ 
  $a_1$ & -0.020 & -0.021 & -0.020 & -0.020 & -0.020 & -0.075 & -0.065 & -0.056 & -0.050 \\ 
  $\alpha_0$ & -1.500 & -1.577 & -1.590 & -1.598 & -1.607 & -1.133 & -1.226 & -1.301 & -1.360 \\ 
  $\tau_0$ & -0.163 & -0.165 & -0.166 & -0.166 & -0.167 & -0.119 & -0.129 & -0.136 & -0.142 \\ \hline
  $K^*=100$ & &  &  &  &  &  &  &  &  \\
  $a_0$ & -1.000 & -1.045 & -1.048 & -1.040 & -1.034 & 0.706 & 0.356 & 0.060 & -0.173 \\ 
  $a_1$ & -0.020 & -0.021 & -0.021 & -0.021 & -0.021 & -0.073 & -0.063 & -0.054 & -0.047 \\ 
  $\alpha_0$ & -1.500 & -1.563 & -1.566 & -1.566 & -1.566 & -1.127 & -1.220 & -1.297 & -1.356 \\ 
  $\tau_0$ & -0.163 & -0.167 & -0.167 & -0.167 & -0.167 & -0.121 & -0.131 & -0.139 & -0.145 \\ \hline
  $K^*=200$ & &  &  &  &  &  &  &  &  \\
  $a_0$ & -1.000 & -0.961 & -0.976 & -0.962 & -0.957 & 0.782 & 0.407 & 0.114 & -0.136 \\ 
  $a_1$ & -0.020 & -0.021 & -0.021 & -0.021 & -0.021 & -0.074 & -0.063 & -0.054 & -0.046 \\ 
  $\alpha_0$ & -1.500 & -1.491 & -1.497 & -1.491 & -1.490 & -1.058 & -1.156 & -1.226 & -1.287 \\ 
  $\tau_0$ & -0.163 & -0.161 & -0.161 & -0.161 & -0.160 & -0.115 & -0.126 & -0.133 & -0.139  \\ \hline \hline
 \multicolumn{2}{|c|}{Gamma lifetimes}& \multicolumn{4}{c|}{Non-contaminated data} &  \multicolumn{4}{c|}{Contaminated data} \\   \hline
    & & & \multicolumn{3}{c|}{$\beta$} & & \multicolumn{3}{c|}{$\beta$}\\ \cline{4-6} \cline{8-10}
  Parameter & True value &QMLE & 0.2 & 0.4 & 0.6 & QMLE & 0.2 & 0.4 & 0.6  \\  \hline
  $K^*=50$ & &  &  &  &  &  &  &  &  \\
  $a_0$ & -1.000 & -1.002 & -1.031 & -1.042 & -1.062 & 1.773 & 1.721 & 1.615 & 1.453 \\ 
  $a_1$ & -0.020 & -0.022 & -0.021 & -0.021 & -0.021 & -0.105 & -0.104 & -0.101 & -0.097 \\ 
  $\alpha_0$ & -1.500 & -1.548 & -1.558 & -1.566 & -1.578 & -0.850 & -0.875 & -0.913 & -0.965 \\ 
  $\tau_0$ & -0.163 & -0.162 & -0.163 & -0.163 & -0.164 & -0.090 & -0.093 & -0.096 & -0.101 \\ \hline
  $K^*=100$ & &  &  &  &  &  &  &  &  \\
  $a_0$ & -1.000 & -1.145 & -1.155 & -1.154 & -1.153 & 1.595 & 1.538 & 1.413 & 1.227 \\ 
  $a_1$ & -0.020 & -0.018 & -0.017 & -0.017 & -0.017 & -0.100 & -0.098 & -0.095 & -0.089 \\ 
  $\alpha_0$ & -1.500 & -1.583 & -1.584 & -1.584 & -1.584 & -0.893 & -0.916 & -0.953 & -1.003 \\ 
  $\tau_0$ & -0.163 & -0.169 & -0.169 & -0.168 & -0.168 & -0.097 & -0.099 & -0.103 & -0.108 \\ \hline
  $K^*=200$ & &  &  &  &  &  &  &  &  \\
  $a_0$ & -1.000 & -0.969 & -0.992 & -0.981 & -0.975 & 1.750 & 1.673 & 1.563 & 1.378 \\ 
  $a_1$ & -0.020 & -0.021 & -0.020 & -0.021 & -0.021 & -0.102 & -0.101 & -0.097 & -0.092 \\ 
  $\alpha_0$ & -1.500 & -1.495 & -1.503 & -1.499 & -1.497 & -0.813 & -0.844 & -0.873 & -0.920 \\ 
  $\tau_0$ & -0.163 & -0.161 & -0.162 & -0.162 & -0.161 & -0.089 & -0.092 & -0.095 & -0.100 \\ 
   \hline
\end{tabular}
\end{table}


\subsubsection{Second simulation scenario: emulating toxicological sacrifice experiments}

We design now a simulation study inspired by toxicological and radiobiological experiments where laboratory animals are sacrificed at scheduled times to assess disease onset. Each subject is inspected only once, resulting in completely censored lifetime data—a setting analogous to one-shot devices. Two correlated failure modes are considered: (i) tumor development and (ii) severe renal damage, both biologically linked through systemic toxicity.

Following the competing risks framework in \cite{Balakrishnan2025}, we assume three stress levels ($x = -2.5, -1, -0.1$) representing increasing exposure intensity, evaluated at four inspection times ($IT = 0.15, 1.3, 2$). For each stress–inspection configuration, $K^*$ animals are tested, with sample sizes $K^* \in \{50, 75, \dots, 250\}$.

Lifetime distributions for both failure modes are generated under two alternative assumptions:

\begin{enumerate}
\item[i)] Weibull distribution with common scale parameter $\beta_j=\exp(s_0+s_1x_j)$ and shape parameter for each failure mode $\eta_{j,g}=\exp(r_{0,g}+r_1x_j)$, with $(s_0,s_1,r_{0,1},r_{0,2},r_1)=(3.5,-0.02,2,2.1,-0.03)$,
\item[ii)] Gamma distribution with common scale parameter $\beta_j=\exp(s_0+s_1x_j)$ and shape parameter for each failure mode $\eta_{j,g}=\exp(r_{0,g}+r_1x_j)$, with $(s_0,s_1,r_{0,1},r_{0,2},r_1)=(-0.7,-0.9,0.9,1,1)$.
\end{enumerate}

The parameters of interest $\boldsymbol{\theta}=(a_0,a_1)^T$ under a) GH copula; b) Frank copula with positive dependence; and c) Frank copula with negative dependence, are taken to be $(-2,-0.02)$, $(1,-0.02)$ and $(-1,+0.02)$, respectively. Positive dependence occurs when one failure mode increases the likelihood of the other, as systemic toxicity can simultaneously trigger tumor development and severe renal damage in the same animal. Negative dependence arises when one failure mode reduces the chance of the other, either by preventing its occurrence or by delaying it; for example, acute renal toxicity may progress so rapidly or severely that it causes early mortality or damages tissue, hindering tumor development.

To evaluate the robustness of the proposed estimators, we contaminate the dataset generated  by artificially changing the observed failures in the last testing condition ($x = -0.1, IT = 2$) as in (\ref{eq:contaminationMC}). The copula parameter $\alpha_0$ and the dependence parameter $\tau_0$ are evaluated under normal operating conditions, with $x_0=-3$.  

\begin{figure}[p]
    \centering
    \includegraphics[width=0.85\linewidth]{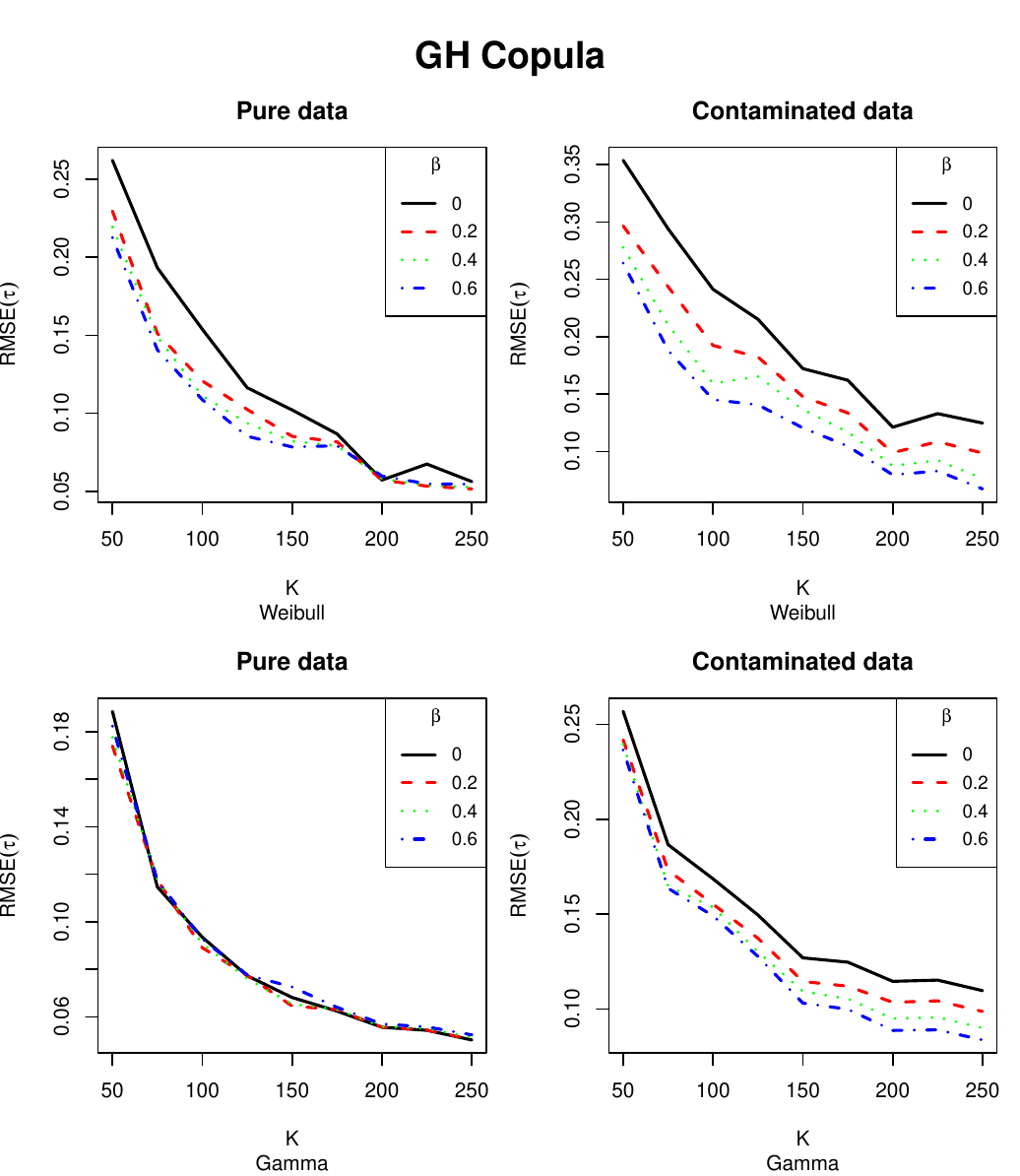}
        \vspace{1cm}
    \caption{RMSE of $\widehat{\tau}_0$  using the GH copula, under varying sample sizes ($K^*\in \{50,75,\dots,250 \}$) and lifetime distributions (Weibull and Gamma). Results are shown for both non-contaminated and contaminated data, with different values of the tuning parameter $\beta$ in the QMDPDE. Estimates are compared with the QMLE ($\beta=0$). \label{fig:GHplot} }
   
\end{figure}

\begin{figure}[p]
    \centering
    \includegraphics[width=0.85\linewidth]{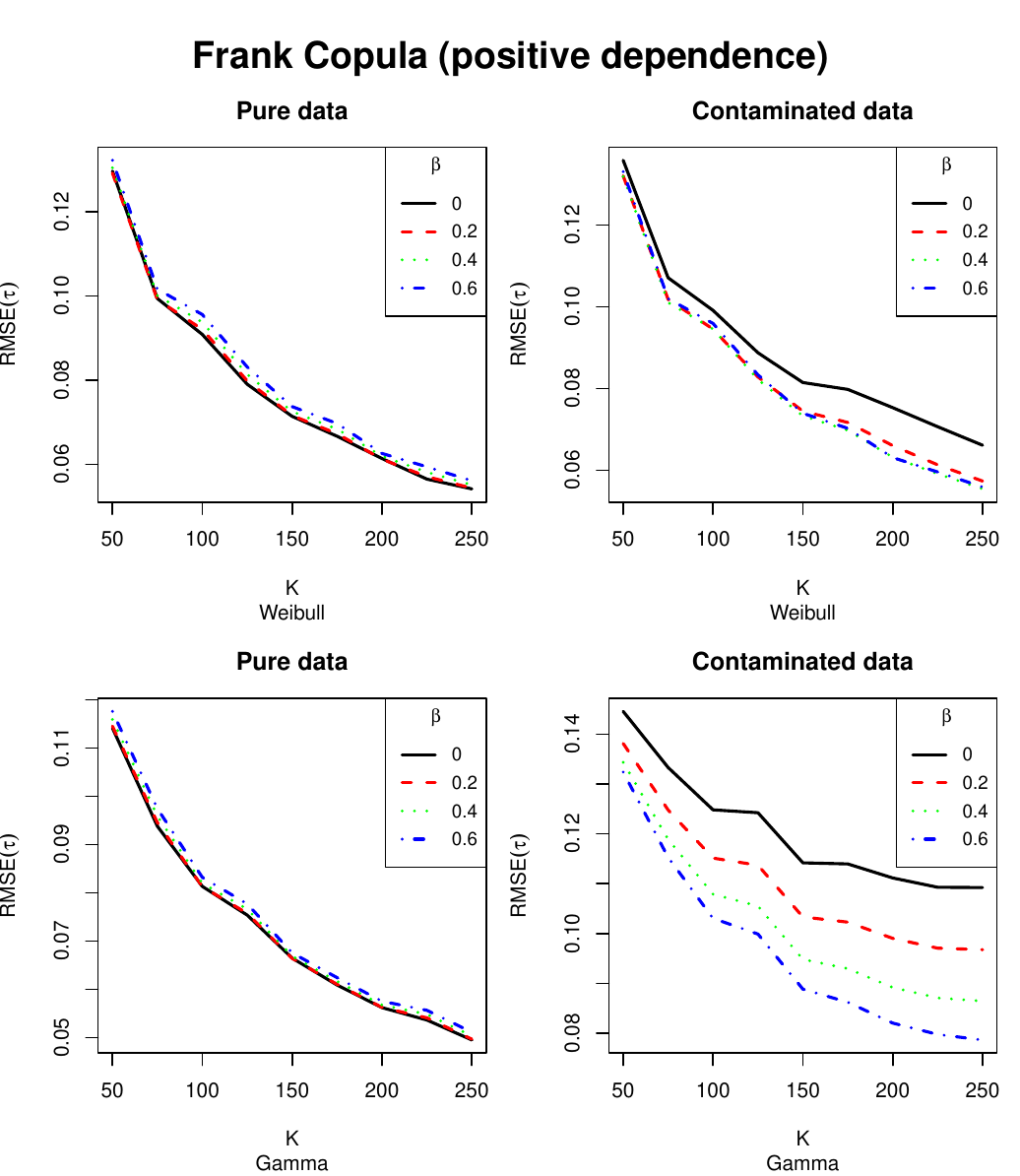}
        \vspace{1cm}
    \caption{RMSE of $\widehat{\tau}_0$  using the Frank copula with positive dependence, under varying sample sizes ($K^*\in \{50,75,\dots,250 \}$) and lifetime distributions (Weibull and Gamma). Results are shown for both non-contaminated and contaminated data, with different values of the tuning parameter $\beta$ in the QMDPDE. Estimates are compared with the QMLE ($\beta=0$). \label{fig:FPplot}}
    
\end{figure}

\begin{figure}[p]
    \centering
    \includegraphics[width=0.85\linewidth]{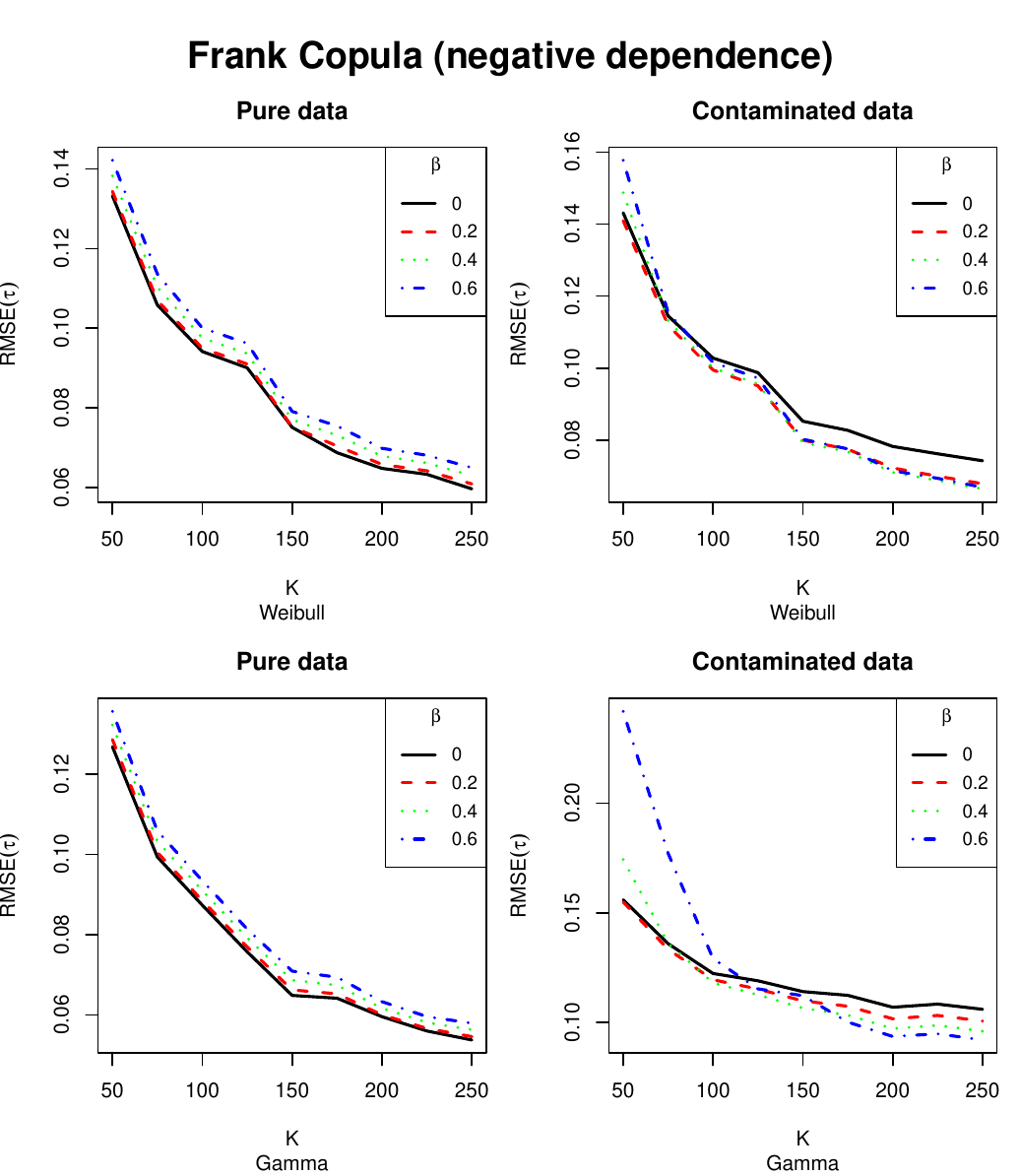}
    \vspace{1cm}
    \caption{RMSE of $\widehat{\tau}_0$  using the Frank copula with negative dependence, under varying sample sizes ($K^*\in \{50,75,\dots,250 \}$) and lifetime distributions (Weibull and Gamma). Results are shown for both non-contaminated and contaminated data, with different values of the tuning parameter $\beta$ in the QMDPDE. Estimates are compared with the QMLE ($\beta=0$).\label{fig:FNplot}}
    
\end{figure}

We obtain the QMDPDE of the model parameters under different tuning parameters $\beta\geq 0$, and compute the Root Mean Square Error (RMSE) of $\widehat{\tau}_0$ for the three copulas. See Figures \ref{fig:GHplot},\ref{fig:FPplot},\ref{fig:FNplot}. While the behavior of the estimators under pure data is less predictable, under contamination  QMDPDEs provide a much robust behavior than QMLE, showing again that these estimators can be a very interesting alternative to QMLE, overall when some contamination is present in our data.

Note that, although the underlying dependence is the same for both marginal models, slight differences may arise due to the variability induced by the marginals in the empirical proportions. This effect is more noticeable for the GH copula near independence, leading to a slightly higher sensitivity of the QMLE under Weibull margins.

\subsection{Real data examples}

\subsubsection{Serial sacrifice data}
We again work with the Serial Sacrifice data presented in Section \ref{sec:SSdata}, a dataset consisting of 704 female mice divided into a control group ($x=0$) and an irradiated group ($x=1$), with each group experiencing two disease categories. We assume both GH and Frank copulas and estimate the QMDPDE of copula parameters for different tuning parameters $\beta \in \{0, 0.2, 0.4, 0.6\}$, as presented in Table \ref{table:Example}. Results for both copulas show a positive correlation between both disease categories, indicating that one disease category may induce another disease category. Note that for the GH copula, there is no a strong evidence of the influence of radiation on this dependence. This is in concordance with findings in  \cite{Ling2021}.

To compare the behavior of different divergences and to select an appropriate one for a given data set, Castilla and Chocano \cite{Castilla2023} proposed minimizing a loss function that relates theoretical and empirical probabilities. Following this approach, we compute the empirical absolute bias (ABias), defined as the absolute difference between the empirical and estimated probabilities, multiplied by 100 to express it as a percentage. We observe that increasing $\beta$ leads to lower values of this estimated error for both copulas, with slightly better results for the Frank copula. This further supports the use of QMDPDE in copula-based models.

\begin{table}[h!!!!]\setlength\tabcolsep{5pt}\def\arraystretch{1.2}
\small
\centering
\caption{Estimates for Serial Sacrifice data. \label{table:Example}}
\vspace{9pt}
\begin{tabular}{|r|rrrr|rrrr|}
\hline
 & \multicolumn{4}{|c|}{GH Copula}& \multicolumn{4}{c|}{Frank Copula}\\
  \hline
  & & \multicolumn{3}{c|}{$\beta$} & & \multicolumn{3}{c|}{$\beta$}\\ \cline{3-5} \cline{7-9}
Parameter & QMLE & 0.2 & 0.4 & 0.6 & QMLE & 0.2 & 0.4 & 0.6 \\ 
  \hline
$a_0$ & -2.135 & -2.146 & -2.154 & -2.156 & 1.342 & 1.264 & 1.224 & 1.185 \\ 
$a_1$ & 0.048 & -0.198 & 0.200 & 0.429 & 0.425 & 0.415 & 0.585 & 0.742 \\ 
$\alpha(x=0)$ & 1.118 & 1.117 & 1.116 & 1.116 & 1.342 & 1.264 & 1.224 & 1.185 \\ 
$\alpha(x=1)$ & 1.124 & 1.096 & 1.142 & 1.178 & 1.767 & 1.679 & 1.809 & 1.927 \\ 
$\tau(x=0)$ & 0.106 & 0.105 & 0.104 & 0.104 & 0.146 & 0.138 & 0.134 & 0.130 \\ 
$\tau(x=1)$ & 0.110 & 0.088 & 0.124 & 0.151 & 0.191 & 0.182 & 0.195 & 0.207 \\ \hline
ABias & 0.899 & 0.913 & 0.891 & 0.874 & 0.691 & 0.690 & 0.674 & 0.668 \\ 
   \hline
\end{tabular}
\end{table}

\subsubsection{Mice tumor toxicological data}
We again work with the mice tumor toxicological data presented in Section \ref{sec:example2}, consisting of 990 mice from both a high-dose 2-AAF group ($x = 150$ ppm) and a control group ($x = 0$ ppm). The outcomes are treated as two distinct failure events—death (I) and tumor development (II). We assume both Gumbel-Hougaard (GH) and Frank copulas and estimate the QMDPDE of the copula parameters for different tuning values $\beta \in {0, 0.2, 0.4, 0.6}$, as presented in Table \ref{table:Example}. Results for both copulas show a positive correlation between the two event categories, suggesting that the occurrence of one may increase the likelihood of the other.

However, dependence between the two events becomes weaker in the high-dose group. At higher doses of 2-AAF, toxic effects may induce mortality through mechanisms unrelated to tumor development, such as systemic organ failure or acute toxicity. This can reduce the joint occurrence of death and tumor, thereby weakening the statistical dependence between these failure modes. In contrast, at lower doses, tumor progression and mortality may share common biological pathways, resulting in stronger dependence. This dose-dependent shift in failure dynamics supports modeling approaches that allow for varying dependence structures across exposure levels.

Finally, we compute the empirical ABias between the empirical and estimated probabilities. We observe that increasing $\beta$ leads to lower values of this estimated error for both copulas,  further supporting the use of QMDPDE in copula-based models.

\begin{table}[h!!!!]\setlength\tabcolsep{5pt}\def\arraystretch{1.2}
\small
\centering
\caption{Parameter estimates for Mice tumor toxicological data. \label{table:ExampleMice}}
\vspace{9pt}
\begin{tabular}{|r|rrrr|rrrr|}
\hline
 & \multicolumn{4}{|c|}{GH Copula}& \multicolumn{4}{c|}{Frank Copula}\\
  \hline
  & & \multicolumn{3}{c|}{$\beta$} & & \multicolumn{3}{c|}{$\beta$}\\ \cline{3-5} \cline{7-9}
Parameter & QMLE & 0.2 & 0.4 & 0.6 & QMLE & 0.2 & 0.4 & 0.6 \\ 
  \hline
 & 0.01 & 0.2 & 0.4 & 0.6 & 0.01 & 0.2 & 0.4 & 0.6 \\ 
  \hline
$a_0$ & 1.323 & 1.585 & 1.999 & 2.794 & 17.994 & 22.719 & 34.870 & 43.294 \\ 
$a_1$ & -0.007 & -0.007 & -0.008 & -0.013 & -0.077 & -0.101 & -0.176 & 0.251 \\ 
$\alpha(x=0)$ & 4.755 & 5.877 & 8.385 & 17.349 & 17.994 & 22.719 & 34.870 & 43.294 \\ 
$\alpha(x=1)$ & 2.358 & 2.816 & 3.093 & 3.233 & 6.390 & 7.588 & 8.438 & 80.995 \\ 
$\tau(x=0)$ & 0.790 & 0.830 & 0.881 & 0.942 & 0.798 & 0.837 & 0.891 & 0.911 \\ 
$\tau(x=1)$ & 0.576 & 0.645 & 0.677 & 0.691 & 0.534 & 0.587 & 0.618 & 0.952 \\ \hline
ABias & 0.926 & 0.838 & 0.768 & 0.739 & 0.939 & 0.859 & 0.783 & 0.650 \\ 
   \hline
\end{tabular}
\end{table}

\section{Discussion \label{sec:conclusions}}

This paper introduces a robust estimation framework for copula models to capture dependence between failure modes in one-shot device testing. The approach is based on divergence measures, which generalize classical MLE. While MLE is efficient under ideal conditions, it is also sensitive to outliers. Through extensive simulations and a real data application, we demonstrate that divergence-based estimators are more robust than MLE, without sacrificing efficiency under regular conditions. The results suggest that divergence-based methods are promising tools for modeling dependent failure mechanisms in one-shot device testing.  

Future research could focus on extending these methods to hierarchical copula models or incorporating covariate information. Finally, applying divergence-based methods to gamma frailty models \cite{Ling2021frailty,Ling2022,Yu2024,Yu2025} represents another promising avenue of investigation.

\paragraph{Funding}{
This work was funded by Universidad Rey Juan Carlos, Project Reference 2025/SOLCON-159637.}

\paragraph{Competing Interests}{
The authors have no conflicts of interest to declare that are relevant to the content of this chapter.}

\section*{Appendix \label{sec:app}}
\addcontentsline{toc}{section}{Appendix}

\subsection*{Some notes on Weibull and Gamma distributions}

In the context of copula-based modeling of one-shot devices with distinct failure modes, we employ Weibull and Gamma distributions for the marginal lifetimes. These distributions provide the necessary flexibility to capture various failure-time behaviors, such as increasing, decreasing, or constant hazard rates. Their analytical tractability, interpretability, and empirical support in reliability engineering make them ideal choices for marginal modeling in the presence of multiple, potentially dependent, causes of failure.

The Weibull distribution with scale parameter $\beta > 0$ and shape parameter $\eta > 0$ has the cumulative distribution function (CDF)
\[
F(t) = 1 - \exp\left[-\left(\frac{t}{\beta}\right)^\eta\right], \quad t \geq 0,
\]
and the corresponding hazard rate function is
\[
h(t) = \frac{\eta}{\beta} \left( \frac{t}{\beta} \right)^{\eta - 1}.
\]
The mean of the Weibull distribution is given by
\[
\mathbb{E}[T] = \beta \, \Gamma\left(1 + \frac{1}{\eta}\right),
\]
where $\Gamma(\cdot)$ denotes the gamma function.

The Gamma distribution with scale parameter $\beta > 0$ and shape parameter $\eta > 0$ has the CDF
\[
F(t) = \frac{1}{\Gamma(\eta)} \, \gamma\left(\eta, \frac{t}{\beta}\right), \quad t \geq 0,
\]
where $\gamma(\cdot, \cdot)$ is the lower incomplete gamma function:
\[
\gamma(\eta, x) = \int_0^x u^{\eta - 1} e^{-u} \, du.
\]
While the hazard rate for the Gamma distribution does not have a simple closed form, it is known to be increasing for shape parameters $\eta > 1$ and decreasing for $\eta < 1$. The mean of the Gamma distribution is:
\[
\mathbb{E}[T] = \eta \beta.
\]

\bigskip

In both distributions, the shape parameter $\eta$ controls the form of the distribution (including the behavior of the hazard rate), while the scale parameter $\beta$ stretches or compresses the distribution along the time axis. This separation of effects makes both distributions highly interpretable in reliability applications.


\begin{thebibliography}{99}

\bibitem{Balakrishnan2021} Balakrishnan, N., Ling, M.H., So, H.Y.: \textit{Accelerated Life Testing of One-shot Devices: Data Collection and Analysis}. John Wiley \& Sons, Hoboken (2021)

\bibitem{Balakrishnan2025} Balakrishnan, N., Castilla, E.: \textit{Statistical Modeling and Robust Inference for One-shot Devices}. Academic Press, Elsevier, San Diego (2025)

\bibitem{Balakrishnan2012} Balakrishnan, N.,  Ling, M.H.:  EM algorithm for one-shot device testing under the exponential distribution. Comput. Stat. Data Anal. \textbf{56}(3), 502-509 (2012)

\bibitem{Balakrishnan2013} Balakrishnan, N., Ling, M.H.: Expectation maximization algorithm for one-shot device accelerated life testing with Weibull lifetimes, and variable parameters over stress. IEEE Trans. Reliab. \textbf{62}(2), 537--551 (2013)

\bibitem{Balakrishnan2014} Balakrishnan, N., Ling, M.H.: Gamma lifetimes and one-shot device testing analysis. Reliab. Eng. Syst. Saf. \textbf{126}, 54--64 (2014)

\bibitem{Balakrishnan2022}  Balakrishnan, N.,  Castilla, E. EM‐based likelihood inference for one‐shot device test data under log‐normal lifetimes and the optimal design of a CSALT plan. Qual. Reliab. Eng. Int. \textbf{38}(2), 780--799 (2022)

\bibitem{Ling2019}  Ling, M. H.: Optimal design of simple step-stress accelerated life tests for one-shot devices under exponential distributions.Probab. Eng. Inform. Sci., \textbf{33}(1), 121-135 (2019)

\bibitem{Tung2025} Tung, H.P.,  Ling, M.H.: Unified Accelerated Life Testing for One-Shot Devices With Weibull Lifetime Distributions. IEEE Trans. Reliab., \textbf{75}, 171-180 (2025).

\bibitem{Ling2026} Ling, M.H., Lin, C.T., Balakrishnan, N.: Acceptance sampling plan under step stress accelerated life test for one-shot devices. Reliab. Eng. Syst. Saf., 112481 (2026).



\bibitem{Zhu2021} Zhu, X., Liu, K., He, M.,  Balakrishnan, N.: Reliability estimation for one-shot devices under cyclic accelerated life-testing. Reliab. Eng. Syst. Saf., \textbf{212}, 107595  (2021) 

\bibitem{Zhu2022}  Zhu, X., Liu, K. Reliability of one-shot device with generalized gamma lifetime under cyclic accelerated life-test. Proc. Inst. Mech. Eng., Part O: J. Risk Reliab., \textbf{236}(6), 1007-1023 (2022)

\bibitem{Zhang2025} Zhang, W., Zhu, X., He, M.,  Balakrishnan, N.: Iterative Regression Algorithm for Parameter Estimation for Nondestructive One-Shot Devices Under Cyclic Accelerated Life Test With Adaptive Proportion of Failure Design. IEEE Trans. Reliab., \textbf{74}(4), 4692 - 4703 (2025).



\bibitem{Balakrishnan2015a} Balakrishnan, N., So, H.Y., Ling, M.H.: EM algorithm for one-shot device testing with competing risks under Weibull distribution. IEEE Trans. Reliab. \textbf{65}(2), 973--991 (2015)

\bibitem{Balakrishnan2015b} Balakrishnan, N., So, H.Y., Ling, M.H.: EM algorithm for one-shot device testing with competing risks under exponential distribution. Reliab. Eng. Syst. Saf. \textbf{137}, 129--140 (2015)

\bibitem{Balakrishnan2024} Balakrishnan, N., Castilla, E.: Robust inference for destructive one-shot device test data under Weibull lifetimes and competing risks. J. Comput. Appl. Math. \textbf{437}, 115452 (2024)

\bibitem{OJ2024} Ortega-Jiménez, P., Pellerey, F., Sordo, M.A., Suárez-Llorens, A.: Probability equivalent level for CoVaR and VaR. Insur. Math. Econ. \textbf{115}, 22--35 (2024)

\bibitem{Capaldo2024} Capaldo, M., Di Crescenzo, A., Pellerey, F.: Generalized Gini’s mean difference through distortions and copulas, and related minimizing problems. Stat. Probab. Lett. \textbf{206}, 109981 (2024)

\bibitem{Capaldo2025a} Capaldo, M., Di Crescenzo, A., Pellerey, F.: Mean distances and dependence structures for lifetimes of systems with shared components. Appl. Stoch. Models Bus. Ind. \textbf{41}(2), e70002 (2025)

\bibitem{Capaldo2025b} Capaldo, M., Navarro, J.: New multivariate Gini’s indices. J. Multivar. Anal. \textbf{206}, 105394 (2025)

\bibitem{Ling2021} Ling, M.H., Chan, P.S., Ng, H.K.T., Balakrishnan, N.: Copula models for one-shot device testing data with correlated failure modes. Commun. Stat. Theory Methods \textbf{50}(16), 3875--3888 (2021)

\bibitem{Ashkamini2023} Ashkamini, Sharma, R., Upadhyay, S.K.: Bayes analysis of one-shot device testing data with correlated failure modes using copula models. Commun. Stat. Simul. Comput. 1--20 (2023)

\bibitem{Salem2024} Salem, M., Salah, R.N.: Bayesian inference for one-shot devices with Weibull dependent failure modes using copulas. J. Indian Soc. Probab. Stat. 1--23 (2024)

\bibitem{Prajapati2023} Prajapati, D., Ling, M.H., Chan, P.S., Kundu, D.: Misspecification of copula for one-shot devices under constant stress accelerated life-tests. Proc. Inst. Mech. Eng. Part O J. Risk Reliab. \textbf{237}(4), 725--740 (2023)

\bibitem{Balakrishnan2019ieee} Balakrishnan, N., Castilla, E., Martín N.,  Pardo, L.: Robust estimators and test-statistics for one-shot device testing under the exponential distribution. IEEE Trans. Inf. Theory, \textbf{65}(5),  3080-3096 (2019)


\bibitem{Balakrishnan2023} Balakrishnan, N., Castilla, E.: Robust estimation based on one-shot device test data under log-normal lifetimes. Statistics. \textbf{57}(5), 1061--1086 (2023)

\bibitem{Baghel2024} Baghel, S.,  Mondal, S.: Analysis of one-shot device testing data under logistic-exponential lifetime distribution with an application to SEER gallbladder cancer data. Appl. Math. Model., \textbf{126}, 159-184 (2024)

\bibitem{Berlin1979} Berlin, B., Brodsky, J., Clifford, P.: Testing disease dependence in survival experiments with serial sacrifice. J. Am. Stat. Assoc. \textbf{74}(365), 5--14 (1979)


\bibitem{Lindsey1993} Lindsey, J.C.,  Ryan, L.M.A three-state multiplicative model for rodent
tumorigenicity experiments. J. R. Stat. Soc., C: Appl. Stat., \textbf{42}(2), 283-300 (1993)


\bibitem{Nelsen2006} Nelsen, R.B.: \textit{An Introduction to Copulas}, 2nd edn. Springer, New York (2006)

\bibitem{Basu1998} Basu, A., Harris, I.R., Hjort, N.L., Jones, M.C.: Robust and efficient estimation by minimising a density power divergence. Biometrika \textbf{85}(3), 549--559 (1998)

\bibitem{Nelson2009} Nelson, W.B.: \textit{Accelerated Testing: Statistical Models, Test Plans, and Data Analysis}, 2nd edn. John Wiley \& Sons, Hoboken (2009)


\bibitem{Castilla2023} Castilla, E., Chocano, P.J.: On the choice of the optimal tuning parameter in robust one-shot device testing analysis. In: Balakrishnan, N. et al. (eds.) \textit{Trends in Mathematical, Information and Data Sciences}. Stud. Syst. Decis. Control, vol. 445, pp. 169--180. Springer, Cham (2023)



\bibitem{Ling2021frailty} Ling, M.H., Balakrishnan, N., Yu, C.,  So, H.Y.: Inference for one-shot devices with dependent k-out-of-M structured components under gamma frailty. Mathematics, \textbf{9}(23), 3032. (2021)

\bibitem{Ling2022}  Ling, M.H.: Optimal constant-stress accelerated life test plans for one-shot devices with components having exponential lifetimes under gamma frailty models. Mathematics, \textbf{10}(5), 840  (2022).

\bibitem{Yu2024} Yu, C.: \textit{Inference for Gamma Frailty Models Based on One-shot Device Data}. PhD thesis, McMaster University (2024). Available at McMaster University’s Institutional Repository

\bibitem{Yu2025} Yu, C., So, H.Y.,  Balakrishnan, N. (2025). Inference for one-shot devices with Weibull component lifetimes under gamma frailty model. In \textit{Stochastic Modeling and Statistical Methods} (pp. 1-34). Academic Press.

\end{thebibliography}
\end{document}